\documentclass[11pt]{article}
\pdfoutput=1
\usepackage{putex}
\usepackage[vcentermath]{youngtab}
\usepackage{subfig}
\usepackage{lscape}

\usepackage{graphicx}
\usepackage{epstopdf}
\usepackage{enumerate}
\usepackage{cite}
\usepackage{tensor}
\usepackage{slashed}
\usepackage{amsmath}
\usepackage{amssymb}
\usepackage{mathrsfs}
\usepackage{lgrind}

\usepackage{bbm}

\usepackage{hyperref}

\numberwithin{equation}{section}

\newcommand {\be} {\begin {equation}}
\newcommand {\ee} {\end {equation}}

\newcommand {\bes} {\begin {equation*}}
\newcommand {\ees} {\end {equation*}}

\newcommand{\eps}{\epsilon}

\def\CH{{\cal H}}

\newcommand{\beq}{\begin{equation}}
\newcommand{\eeq}{\end{equation}}

\def\be{ \begin{equation} }
\def\ee{ \end{equation} }

\begin{document}

\preprint{PUPT-2481}

\institution{PU}{Department of Physics, Princeton University, Princeton, NJ 08544}
\institution{PCTS}{Princeton Center for Theoretical Science, Princeton University, Princeton, NJ 08544}

\title{
Generalized $F$-Theorem and the $\epsilon$ Expansion
}

\authors{Lin Fei,\worksat{\PU} Simone Giombi,\worksat{\PU} Igor R.~Klebanov\worksat{\PU,\PCTS} and Grigory Tarnopolsky\worksat{\PU}
}

\abstract{Some known constraints on Renormalization Group flow take the form of inequalities: in even dimensions they refer to the coefficient $a$ of the Weyl anomaly,
while in odd dimensions to the sphere free energy $F$. In recent work \cite{Giombi:2014xxa} it was suggested that the $a$- and $F$-theorems may be viewed as special cases of a
Generalized $F$-Theorem valid in continuous dimension. This conjecture states that, for any RG flow from one conformal fixed point to another, $\tilde F_{\rm UV} > \tilde F_{\rm IR}$, where $\tilde F=\sin (\pi d/2)\log Z_{S^d}$.
 Here we provide additional evidence in favor of the Generalized $F$-Theorem. We show that it holds in conformal perturbation theory, i.e.
for RG flows produced by weakly relevant operators.
We also study a specific example of the Wilson-Fisher $O(N)$ model and define this CFT on the sphere $S^{4-\epsilon}$, paying careful attention to the beta functions
for the coefficients of curvature terms. This allows us to develop the $\epsilon$ expansion of $\tilde F$ up to order $\epsilon^5$.
Pad\' e extrapolation of this series to $d=3$ gives results that are around $2-3\%$ below the free field values for small $N$.
We also study RG flows which include an anisotropic perturbation breaking the $O(N)$ symmetry; we again find that the results are
consistent with $\tilde F_{\rm UV} > \tilde F_{\rm IR}$.
}

\date{}
\maketitle

\tableofcontents

\section{Introduction and Summary}

A well-known set of constraints on Renormalization Group in $d$-dimensional relativistic Quantum Field Theory (QFT) takes the form of inequalities:
if an RG trajectory leads from a short-distance unitary Conformal Field Theory (CFT)
to a long-distance one, then a certain positive quantity defined on the space of CFTs decreases.
 In even dimensions $d$, using the modern terminology such theorems may be called the $a$-theorems.
The Weyl anomaly equation takes the form
\be
\langle T^\mu_\mu \rangle \sim (-1)^{d/2} a E_d + \sum_i c_i I_i\,,
\ee
where $E_d$ is the Euler density term, which is present in all even $d\geq 2$, and $c_i$ are the coefficients of other Weyl invariant curvature terms.
The universal Weyl anomaly coefficient $a$ may be extracted from the free energy on a sphere of radius $R$: $F=-\log Z_{S^d}=(-1)^{d/2} a \log R$.
In even $d$, the RG inequalities take the form \cite{Zamolodchikov:1986gt,Cardy:1988cwa,Komargodski:2011vj,Komargodski:2011xv,Cordova:2015vwa,Cordova:2015fha}
\be
\label{ath}
a_{\rm UV} > a_{\rm IR}\ .
\ee
 In $d=2$ this is well-known as the $c$-theorem \cite{Zamolodchikov:1986gt}, because there is only one Weyl anomaly coefficient, and it
is standard to define $c=3 a$.
In $d=4$ a non-perturbative
proof of (\ref{ath}) was found in \cite{Komargodski:2011vj,Komargodski:2011xv}, long after the early work of \cite{Cardy:1988cwa,Jack:1990eb}.
Very recently \cite{Cordova:2015vwa,Cordova:2015fha}, there was major progress towards establishing (\ref{ath}) for supersymmetric flows in $d=6$, building on the earlier work including \cite{Maxfield:2012aw,Elvang:2012st}.

In odd dimensions the situation is quite different because there are no Weyl anomalies. This actually has some advantages: for a CFT, the sphere
free energy $F=-\log Z_{S^d}$ is a finite, radius independent quantity, which has no ambiguities because there are no Weyl invariant terms constructed purely out of the curvature tensor.
In $d=3$ it was conjectured \cite{Jafferis:2011zi} that the RG inequality takes the form $F_{\rm UV} > F_{\rm IR}$. This $F$-theorem can be equivalently formulated in terms of the
entanglement entropy across a circle \cite{Myers:2010xs,Casini:2011kv}.
A proof of the three-dimensional $F$-theorem has been found using properties of the entanglement entropy in relativistic theories \cite{Casini:2012ei} (see also \cite{Liu:2012ee}). For other odd dimensions, it was conjectured \cite{Klebanov:2011gs} that the RG inequality takes the form $\tilde F_{\rm UV} > \tilde F_{\rm IR}$, where
$\tilde F=(-1)^{(d-1)/2} \log Z_{S^d}$.\footnote{In the special case $d=1$ this conicides with the $g$-theorem
for boundary conformal field theory \cite{Affleck:1991tk}.} Supporting evidence for this conjecture included
a calculation of the change in $\tilde F$ for the RG flow produced by a weakly relevant operator of dimension $d-\epsilon$ \cite{Klebanov:2011gs}. Such calculations, as well as their analogues
in even dimensions which apply to the change in $a$ \cite{Zamolodchikov:1986gt, Cardy:1988cwa,Komargodski:2011xv}, can be carried out perturbatively in $\epsilon$.

The similarity between the RG inequalities in even and odd dimensions (both of them can be phrased in terms of the free energy on $S^d$ or, equivalently, in terms of
the entanglement entropy across $S^{d-2}$ \cite{Myers:2010xs}) has led to the idea that they are special cases of the RG inequality valid in continuous dimension \cite{Giombi:2014xxa}:\footnote{Introduction of the factor $\sin (\pi d/2)$ is just a convenient choice; a further multiplication of $\tilde F$ by a positive function of
$d$ does not change the conjectured inequality.}
\be
\label{GenF}
\tilde F_{\rm UV} > \tilde F_{\rm IR}\ , \qquad \tilde F=\sin (\pi d/2)\log Z_{S^d}=- \sin (\pi d/2) F\ .
\ee
In even dimensions, the factor $\sin (\pi d/2)$ cancels the pole present in $F$, and we have $\tilde F= \pi a/2$; therefore (\ref{GenF}) reduces to the $a$-theorem.
In odd dimensions, this definition reduces to that proposed in \cite{Klebanov:2011gs}. Thus, the ``Generalized $F$-Theorem" (\ref{GenF}) smoothly interpolates
between the $a$-theorems in even $d$ and the $F$-theorems in odd $d$.

In \cite{Giombi:2014xxa} several pieces of evidence were provided in favor of
 the Generalized $F$-Theorem.\footnote{Some holographic evidence for the
 Generalized $F$-Theorem was also provided in \cite{Kawano:2014moa}.} In free conformal scalar and fermion theories, $\tilde F$ is positive for all $d$.
 For example, for a conformally coupled scalar it is
 \be
 \tilde F_s(d) =\frac{1}{\Gamma\left(1+d\right)}\int_0^1 du\, u\sin\pi u\, \Gamma\left(\frac{d}{2}+u\right)\Gamma\left(\frac{d}{2}-u\right)\ , \label{tFfree}
 \ee
 which is a smooth, monotonically decreasing function of $d$.
   Therefore, (\ref{GenF}) is valid for the RG flows produced by the
 scalar or fermion mass terms -- such flows lead to trivial theories where $\tilde F=0$.
  Other tractable RG flows include those produced by double-trace operators in large $N$ theories. As is well-known \cite{Witten:2001ua,Gubser:2002vv},
 turning on a relevant perturbation $O_{\Delta}^2$, which is
 the square of
 of a primary scalar operator of dimension $\Delta<d/2$, makes a large $N$ CFT flow to another CFT where the corresponding operator has dimension
 $d-\Delta+{\cal O}(1/N)$. This produces the following change in $\tilde F$ \cite{Diaz:2007an,Giombi:2014xxa}:
 \begin{equation}
\tilde F_{\rm IR}- \tilde F_{\rm UV} = \frac{1}{\Gamma\left(1+d\right)}\int_0^{\Delta-\frac{d}{2}} du\, u\sin\pi u\, \Gamma\left(\frac{d}{2}+u\right)\Gamma\left(\frac{d}{2}-u\right) + {\cal O}(1/N) \,.
\label{dtF}
\end{equation}
This is negative in the entire range $(d-2)/2 < \Delta < d/2$ where the UV and IR CFTs are unitary, establishing consistency with the Generalized $F$-Theorem. However, when $\Delta$ is sufficiently
far below the unitarity bound $(d-2)/2$, then the inequality (\ref{GenF}) is violated (for $d=3$ this was discussed in \cite{Giombi:2013yva}).
This provides a simple explicit example of how (\ref{GenF}) may be violated in non-unitary theories.\footnote{
 In non-integer dimensions all theories appear to exhibit a subtle form of non-unitarity \cite{Hogervorst:2014rta}, but hopefully this does not invalidate
 the Generalized $F$-Theorem for theories that are dimensional continuations of unitary theories in integer $d$.}

Explicit applications of the $a$-theorem in $d>2$ have been mostly to supersymmetric RG flows
(see, for instance, \cite{Anselmi:1997am,Intriligator:2003jj,Kutasov:2003iy,Cordova:2015vwa,Cordova:2015fha,Heckman:2015ola,Heckman:2015axa}), because few interacting
non-supersymmetric CFTs are known in $d=4$ and none in $d=6$. On the other hand, in $2n-\eps$ dimensions there is a multitude of non-supersymmetric
RG flows connecting perturbative fixed points \cite{Wilson:1971dc,Hasenfratz:1991it,ZinnJustin:1991yn,Halperin:1973jh,Fei:2014yja,Fei:2014xta,Gracey:2015tta}. This
opens the possibility of many tests of the Generalized $F$-Theorem in $2n-\eps$ dimensions, and some of them were carried out in \cite{Giombi:2014xxa}.
In this paper we extend these calculations, focusing mostly on
the Wilson-Fisher fixed point in $4-\epsilon$ dimensions \cite{Wilson:1971dc}, which is the
$O(N)$ symmetric theory of $N$ real scalar fields $\phi^i$, $i=1, \ldots, N$,
with interaction ${\lambda\over 4} (\phi^i \phi^i)^2$. For large $N$, this is a double-trace operator with $\Delta=d-2$. Furthermore,
the fact that the coupling constant at the IR fixed point is of order $\epsilon$ allows one to develop the $\epsilon$ expansions for
the critical exponents \cite{Wilson:1971dc}; this works well for all values of $N$ including $N=1$, i.e.
for the Ising model \cite{Wilson:1973jj}. Fortunately, $\tilde F$ is also amenable to
$\epsilon$ expansion using renormalization of the $O(N)$ model on the sphere $S^{4-\epsilon}$. In  \cite{Giombi:2014xxa} the contributions of interactions to $\tilde F$ were determined
up to  ${\cal O}(\epsilon^4)$. The leading interaction correction, which is of order $\epsilon^3$, did not require renormalization and was quite straightforward.
However, at the next order one needs to renormalize the theory including the effects of the terms quadratic in the curvature.
In section \ref{WF-curved} we elucidate the definition of the Wilson-Fisher fixed point on $S^{4-\epsilon}$ following
\cite{Drummond:1977dg,Brown:1980qq, Hathrell:1981zb,Jack:1983sk,Jack:1990eb}.
We find that it is important to define the IR theory by setting {\it all} the beta functions to zero, including the beta functions for the coefficients of the curvature terms.
Applying this procedure up to order $\eps^5$ we find
\begin{align}
&\tilde{F}_{O(N)}=N \tilde{F}_s(4-\eps)-\frac{\pi  N (N+2) }{576 (N+8)^2}\epsilon ^3
 -\frac{\pi  N (N+2) (13 N^{2}+370N+1588) }{6912 (N+8)^4}\epsilon^4\notag\\
 &+\frac{\pi  N(N+2)}{414720 (N+8)^6}\left(10368 (N+8) (5 N+22) \zeta (3)-647 N^4-32152 N^3-606576 N^2-3939520N\right.\notag\\
 &~~~~~~~~~~~~~~\qquad\left.+30 \pi ^2 (N+8)^4-8451008\right)\eps^5+\mathcal{O}(\eps^6)\, .
 \label{tFON}
\end{align}
As a byproduct of our analysis of the $O(N)$ model on $S^{4-\epsilon}$, we
extend the previous results \cite{Brown:1980qq, Hathrell:1981zb,Jack:1983sk,Jack:1990eb} and determine the beta function for the Euler density term to order $\lambda^5$:
\begin{equation}
\beta_b= \epsilon b -\frac{N}{360(4\pi)^2}-\frac{ N (N+2) (3N+14)}{48 (4 \pi )^{10}}\lambda^4 +
\frac{N(N+2) \big(3(24+7N)+4(5 N+22) \zeta (3)\big)}{15 (4 \pi )^{12}}\lambda^5+
\mathcal{O}(\lambda^6)\, .
\label{betaEuler}
\end{equation}
As far as we know, the $\mathcal{O}(\lambda^5)$ term
has not appeared in the previous literature.
 We also present a calculation of the sphere free energy and curvature beta functions in the most general
quartic scalar field theory in Appendix \ref{gen-coup}.

A constrained Pad\' e approximation of the series (\ref{tFON}) for $N=1$, using the boundary condition that the 2-d Ising model has $\tilde F=\pi/12$, corresponding to $c=1/2$,
gives that in 3-dimensions $F_{\rm Ising}/F_s \approx 0.976$. This is consistent with the $F$-theorem, but the surprise is how close $F_{\rm Ising}$ is to the free field value.\footnote{The numbers
we get from the Pad\' e approximants are close to the the estimates in \cite{Giombi:2014xxa} which were obtained without the use of resummation.}
For comparison, we note that the scaling dimension of $\phi^i$, $\Delta_\phi\approx 0.5182$ is around $3.6 \%$ above the free field value.
The coefficient of the stress tensor 2-point function, $c_T$, is also known to be close to the free field value for the 3-d Ising model:
$c_T^{\rm 3d\,Ising}/c_T^{s}\approx 0.9466$ \cite{ElShowk:2012ht,El-Showk:2014dwa}.

In section \ref{confpert} we also provide a check of the Generalized $F$-Theorem for weakly relevant perturbations. When an RG flow in $d$ dimensions is sourced by an operator $O$ of
dimension $d-\epsilon$ then there is a nearby IR fixed point, and the change in $\tilde F$ is
\begin{equation}
\tilde F_{\textrm{IR}}-\tilde F_{\textrm{UV}}=-\frac{\pi \Gamma\left(\frac{d}{2}\right)^2}{\Gamma\left(1+d\right)}
\frac{{\cal C}_2^3}{3 {\cal C}_3^2}\epsilon^3+\mathcal{O}(\eps^4)\, ,
\label{deltaFconfpert}
\end{equation}
where ${\cal C}_2$ and ${\cal C}_3$ are the coefficients of the two- and three-point functions of $O$ in the UV CFT, (\ref{2pt3pt}).
In odd $d$ this result agrees with \cite{Klebanov:2011gs},
and in even $d$ with the change in $a$-anomaly computed in \cite{Komargodski:2011xv}.
For a unitary CFT, ${\cal C}_2>0$ and ${\cal C}_3$ is real. So we find that the Generalized $F$-Theorem holds to leading
order in conformal perturbation theory for all $d$. We have also extended the conformal perturbation theory
approach to determine the term of order $\eps^4$ in specific models.
For example, for the $O(N)$ model it reproduces this term in (\ref{tFON}).

In section \ref{confseveral} we generalize the conformal perturbation approach to the case of several weakly relevant operators.
We obtain a concise formula (\ref{deltaFMany}) for the leading order change in $\tilde F$. As a specific application, we study the
$O(N)$ model with an extra ``cubic anisotropy" operator $\sum_i \phi_i^4$ which breaks the $O(N)$ symmetry. The conformal perturbation approach allows us to calculate
the $\tilde F$ to order $\eps^3$ at the different fixed points of this theory with two coupling constants. The results are found to be in agreement with the direct approach of section \ref{WF-curved} which uses renormalization of the theory on $S^{4-\eps}$.


\section{The Wilson-Fisher fixed points in curved space}
\label{WF-curved}
The renormalization of interacting scalar field theory in $d=4-\epsilon$ on a curved space was studied in \cite{Brown:1980qq, Hathrell:1981zb} for
the single scalar theory, and in \cite{Jack:1983sk, Jack:1990eb} for the more general multicomponent theory (see Appendix \ref{gen-coup}). In the case of $N$ real scalar fields
with $O(N)$ invariant interaction, the full bare action on a curved manifold is
\begin{align}
S = \int d^{d}x \sqrt{g}\left(\frac{1}{2}\big((\partial_{\mu}\phi^{i}_{0})^{2}+ \frac{d-2}{4(d-1)}\mathcal{R}(\phi_{0}^{i})^{2}\big)+\frac{\lambda_{0}}{4}(\phi_{0}^{i}\phi_{0}^{i})^{2}+\frac{1}{2}\eta_{0}H(\phi_{0}^{i})^{2}+a_{0}W^2+b_{0}E+c_{0}H^{2} \right),
\label{bare-S}
\end{align}
where  $\mathcal{R}$ is the Ricci scalar, $H=\frac{ \mathcal{R}}{d-1}$, and
\begin{align}
&W^2= \mathcal{R}_{\mu\nu\lambda\rho}  \mathcal{R}^{\mu\nu\lambda \rho} - \frac{4}{d-2}  \mathcal{R}_{\mu\nu} \mathcal{R}^{\mu\nu} + \frac{2}{(d-2)(d-1)} \mathcal{R}^{2},
\notag\\& E=  \mathcal{R}_{\mu\nu\lambda\rho}  \mathcal{R}^{\mu\nu\lambda \rho} -4 \mathcal{R}_{\mu\nu} \mathcal{R}^{\mu\nu} + \mathcal{R}^{2}.
\end{align}
Here $W^2$ is the square of the Weyl tensor, and $E$ is the Gauss-Bonnet term which is the topological Euler density in $d=4$ (it is not topological in $d=4-\epsilon$,
but we will still refer to it as the Euler density in what follows). In (\ref{bare-S}), we have separated an arbitrary coupling $\eta_0$ to the scalar curvature from
the conformal coupling term $\frac{d-2}{4(d-1)}\mathcal{R}$. The action (\ref{bare-S}) contains all possible terms which are marginal in $d=4$, which have to
be included in order to consistently renormalize the theory in $d=4-\epsilon$. Note that in $d=4-\epsilon$ the coupling $\lambda_0$ has dimension $\epsilon$,
$a_0, b_0, c_0$ dimension $-\epsilon$, and $\eta_0$ is dimensionless in any $d$. We have omitted a mass term for the scalar field since we will be interested in the conformal
theory, and the mass can be consistently set to zero in dimensional regularization.

Since the theory is renormalizable, the divergences in the free energy $F=-\log Z$ on a general manifold can be cancelled by expressing the bare couplings
$\lambda_0, a_0, b_0, c_0, \eta_0$ in terms of renormalized ones $\lambda, a, b, c, \eta$. The renormalization of the quartic coupling is well known from
flat space and reads
\begin{equation}
\lambda_0  =\mu^{\epsilon}\bigg(\lambda+\frac{ (N+8)}{8 \pi ^2 \epsilon }\lambda^2+ \left(\frac{(N+8)^2}{64 \pi ^4 \epsilon ^2}-\frac{3 (3 N+14)}{128 \pi ^4 \epsilon }\right)\lambda^{3}+...\bigg)\,.
\label{lam-bare}
\end{equation}
Requiring $\frac{d\lambda_0}{d \log \mu}=0$ this yields the beta function \cite{Kleinert:2001ax}
\begin{equation}
\label{betalam}
\beta_{\lambda} = -\epsilon \lambda +\frac{N+8}{8\pi^2}\lambda^2-\frac{3(3N+14)}{64\pi^4}\lambda^3
+\frac{\left(33 N^2+480 N \zeta (3)+922N+2112 \zeta(3)+2960\right)}{4096 \pi ^6}\lambda^4+\ldots\,.\\
\end{equation}
The renormalization of the curvature couplings takes the form \cite{Brown:1980qq, Hathrell:1981zb}
\begin{equation}
\begin{aligned}
&a_0 =\mu^{-\epsilon}\left(a+L_a \right), \\
&b_0 = \mu^{-\epsilon}\left(b+L_b \right),  \\
&c_0 =\mu^{-\epsilon}\left(c+L_c +\eta L_{\kappa}+\eta^2 L_{\Lambda}\right), \\
&\eta_0 = (\eta+L_{\eta})Z_2^{-1},
\label{curv-ren}
\end{aligned}
\end{equation}
where $a, b, c, \eta$ are dimensionless renormalized parameters, and $L_a, L_b, L_c, L_{\kappa}, L_{\Lambda}, L_{\eta}$ are
functions of $\lambda$ alone with pure poles in $\epsilon$,\footnote{We use a minimal subtraction scheme.} namely
\begin{equation}
L_a = \sum_{i=1}^{\infty} \frac{L_a^{(i)}(\lambda)}{\epsilon^i}\,,\qquad L_b = \sum_{i=1}^{\infty} \frac{L_b^{(i)}(\lambda)}{\epsilon^i}\,,~~~\ldots
\label{L-MS}
\end{equation}
and similarly for $c, \eta, \kappa, \Lambda$. In the case of the quartic $O(N)$ theory (\ref{bare-S}), these functions are given
explicitly by
\begin{equation}
\begin{aligned}
&L_a =\frac{a_{01}}{\epsilon}+\frac{a_{21}}{\epsilon}\lambda^2+\mathcal{O}(\lambda^3)\,,\quad
L_b =\frac{b_{01}}{\epsilon}+\frac{b_{41}}{\epsilon}\lambda^4
+\left(\frac{b_{52}}{\epsilon^2}+\frac{b_{51}}{\epsilon}\right)\lambda^5+\mathcal{O}(\lambda^6)\,,\quad
L_c =\frac{c_{51}}{\epsilon}\lambda^5+\mathcal{O}(\lambda^6)\,,
\label{curv-ren2}
\end{aligned}
\end{equation}
where the values of the known coefficients are \cite{Brown:1980qq, Hathrell:1981zb, Jack:1983sk}
\begin{equation}
\begin{aligned}
&a_{01}=\frac{N}{120(4\pi)^2}\,,~~~~a_{21}=-\frac{N(N+2)}{216(4\pi)^6}\,,\\
&b_{01}=-\frac{N}{360(4\pi)^2}\,,~~~~b_{41}=-\frac{ N (N+2) (3N+14)}{240 (4 \pi )^{10}}\,,~~~~b_{52}=\frac{(N+8)}{12\pi^2} b_{41}\,,\\
&c_{51}=\frac{N (N+2)(N+8) (3N+14)}{360 (4 \pi )^{12}}\,.
\label{pole-coeff}
\end{aligned}
\end{equation}
We will be able to determine the coefficient $b_{51}$ by a direct perturbative calculation on the sphere in the next section, and the result is given in
eq. (\ref{b51-Euler}).\footnote{Note that, when working on the sphere, we cannot distinguish the divergences
associated with the Euler term from those associated with the ${\mathcal R}^2$ term. Therefore, in order to determine $b_{51}$, we have
to assume the result for $c_{51}$ obtained in \cite{Hathrell:1981zb, Jack:1983sk}.}
The functions $L_{\kappa}$ and $L_{\Lambda}$ will not play an important role in what follows, but they have the structure
\begin{equation}
L_{\Lambda}=\frac{N}{2(4\pi)^2}\frac{1}{\epsilon}+\frac{N(N+2)}{(4\pi)^4}\frac{\lambda}{\epsilon^2}+\mathcal{O}(\lambda^2)\,,\qquad L_{\kappa}=\mathcal{O}(\lambda^4)\,.
\end{equation}
Finally, in the $\eta$ renormalization, $Z_2$ is the $Z$-factor in the flat space renormalization of the $\phi^2$ operator, which is related to
its anomalous dimension by \cite{Kleinert:2001ax}
\begin{equation}
\begin{aligned}
&\beta_{\lambda}\frac{\partial }{\partial \lambda}\log Z_2 = \gamma_{\phi^2}=\frac{(N+2)}{8\pi^2}\lambda-\frac{5(N+2)}{128\pi^4}\lambda^2+\mathcal{O}(\lambda^3)
\end{aligned}
\end{equation}
and the function $L_{\eta}$ starts at order $\lambda^4$ \cite{Brown:1980qq, Hathrell:1981zb}
\begin{equation}
L_{\eta} = \eta_{41}\frac{\lambda^4}{\epsilon}+\mathcal{O}(\lambda^6)\,.
\end{equation}
The coefficient $\eta_{41}$ may be read off from \cite{Hathrell:1981zb, Jack:1983sk}, but we will not need its explicit form here.
Note that the $\lambda$ independent terms  in  $L_a$, $L_b$ and $L_{\Lambda}$, which are proportional to $N$, are those required
to cancel the divergencies of the free field determinant,
$F_{\textrm{free}}=\frac{N}{2}\log \det\left(-\nabla^2+\frac{(d-2)}{4(d-1)}\mathcal{R}+\frac{\eta}{d-1} \mathcal{R}\right)$,
which may be extracted for an arbitrary manifold using heat kernel methods.

After expressing the bare couplings in terms of renormalized ones, the free energy is finite in the limit $\epsilon \rightarrow 0$ for any values of
the renormalized parameters $\lambda, a, b, c, \eta$. The renormalization process implies that the curvature parameters $a, b , c, \eta$ acquire
a scale dependence, just like $\lambda$, and have their own beta functions. Requiring that the bare couplings are independent of $\mu$,
from (\ref{curv-ren}) one obtains the beta functions
\begin{equation}
\begin{aligned}
&\beta_a = \mu \frac{da}{d\mu} = \epsilon a +\hat \beta_a(\lambda)\,,\\
&\beta_b = \mu \frac{db}{d\mu}=\epsilon b +\hat\beta_b(\lambda)\,,\\
&\beta_c = \mu \frac{dc}{d\mu}=\epsilon c + \hat \beta_c (\lambda)+\eta \hat \beta_{\kappa}(\lambda)+\eta^2 \hat \beta_{\Lambda}(\lambda)\,,\\
&\beta_{\eta} = \mu \frac{d\eta}{d\mu} = \eta \gamma_{\phi^2}+\hat\beta_{\eta}(\lambda)\,,
\label{all-beta}
\end{aligned}
\end{equation}
where the hatted beta functions depend on $\lambda$ only (and not on $\epsilon$) and are given by
\begin{equation}
\begin{aligned}
&\hat \beta_a = \left(\epsilon -\beta_{\lambda}\frac{\partial }{\partial \lambda}\right)L_a
=(1+\lambda \frac{\partial }{\partial \lambda})L^{(1)}_a=a_{01}+3 a_{21}\lambda^2+\ldots\,, \\
&\hat \beta_b = \left(\epsilon -\beta_{\lambda}\frac{\partial }{\partial \lambda}\right)L_b
=(1+\lambda \frac{\partial }{\partial \lambda})L^{(1)}_b= b_{01}+5 b_{41}\lambda^4+6 b_{51}\lambda^5+\ldots\,, \\
&\hat \beta_c = \left(\epsilon -\beta_{\lambda}\frac{\partial }{\partial \lambda}\right)L_c-\hat \beta_{\eta}L_{\kappa}
=(1+\lambda \frac{\partial }{\partial \lambda})L^{(1)}_c= 6 c_{51}\lambda^5+\ldots\,, \\
&\hat \beta_{\eta} = \left(\gamma_{\phi^2}-\beta_{\lambda}\frac{\partial }{\partial \lambda}\right)L_{\eta}
=\lambda \frac{\partial }{\partial \lambda} L^{(1)}_{\eta}= 4\eta_{41}\lambda^4+\ldots\,, \\
&\hat \beta_{\Lambda} = \left(\epsilon -\beta_{\lambda}\frac{\partial }{\partial \lambda}-2\gamma_{\phi^2}\right)L_{\Lambda} =
(1+\lambda \frac{\partial }{\partial \lambda})L^{(1)}_{\Lambda} =\mathcal{O}(\lambda^0)\,,\\
&\hat \beta_{\kappa} = \left(\epsilon -\beta_{\lambda}\frac{\partial }{\partial \lambda}-\gamma_{\phi^2}\right)L_{\kappa}-2\hat \beta_{\eta}L_{\Lambda}
=(1+\lambda \frac{\partial }{\partial \lambda})L^{(1)}_{\kappa}=\mathcal{O}(\lambda^4)\,,
\label{hat-beta}
\end{aligned}
\end{equation}
where $L_{a}^{(1)}$, $L_b^{(1)}$, ..., denote the residue of the simple pole as defined in (\ref{L-MS}). As usual, the beta functions depend
only on the coefficients of the simple pole, but there is a tower of consistency conditions ('t Hooft identities) relating higher poles to lower ones.

Since the bare couplings are independent of the renormalization scale, the free energy satisfies the identity
\begin{equation}
\mu\frac{d}{d\mu}F(\lambda_0,a_0,b_0,c_0,\eta_0)=0\,,
\end{equation}
which implies that when $F$ is expressed in terms of the renormalized couplings, it obeys the Callan-Symanzik equation
\begin{equation}
\left( \mu \frac{\partial}{\partial\mu}+\beta_{\lambda}\frac{\partial}{\partial \lambda}+
\beta_{\eta} \frac{\partial}{\partial\eta}
+\beta_a \frac{\partial}{\partial a}+
\beta_b \frac{\partial}{\partial b}+
\beta_c \frac{\partial}{\partial c}\right)F=0\,.
\label{CS}
\end{equation}
Note that the dependence of $F$ on the renormalized parameters is simply linear in $a, b, c$, since the corresponding curvature terms are
field independent additive terms, therefore we may split
\begin{equation}
\label{F-split}
F=F_{\phi}(\lambda,\eta,\mu)+ \mu^{-\epsilon} a \int d^d x \sqrt{g} W^2+\mu^{-\epsilon} b \int d^d x \sqrt{g} E + \mu^{-\epsilon} c \int d^d x\sqrt{g} H^2\,,
\end{equation}
where $F_{\phi}(\lambda,\eta,\mu)$ is finite for any $\lambda,\eta$.

So far we have been working on a general curved manifold. Let us now specialize to the case of a round sphere $S^d$ of radius $R$. In this case,
we have $ \mathcal{R}_{\mu\nu\lambda\rho}  \mathcal{R}^{\mu\nu\lambda \rho}=2d(d-1)/R^{4},\; \mathcal{R}_{\mu\nu} \mathcal{R}^{\mu\nu}=d(d-1)^{2}/R^{4}$
and $\mathcal{R}= d(d-1)/R^{4}$, therefore
\begin{align}
W^2=0\,, \qquad E= d(d-1)(d-2)(d-3)/R^{4}\,,\qquad H= d/R^{2}.
\end{align}
On the sphere, the radius dependence of the free energy is constrained to be $F=F(\lambda, a,b,c,\eta,\mu R)$, and so we see that
\begin{equation}
\mu \frac{\partial F}{\partial \mu}=R\frac{dF}{dR}=-\int_{S^d} d^d x \sqrt{g} \langle T^{\mu}_{\mu} \rangle\,,
\end{equation}
where we have used the conventions $T_{\mu\nu} = \frac{2}{\sqrt{g}}\frac{\delta S}{\delta g^{\mu\nu}}$.
Therefore from the Callan-Symanzik equation (\ref{CS}) we can deduce the relation
\begin{equation}
\int_{S^d} d^d x \sqrt{g} \langle T^{\mu}_{\mu} \rangle =  \beta_{\lambda} \frac{\partial F}{\partial \lambda}+\beta_{\eta}\frac{\partial F}{\partial \eta}
+\mu^{-\epsilon}\beta_b \int_{S^d} d^dx \sqrt{g} E +\mu^{-\epsilon} \beta_c \int_{S^d} d^d x \sqrt{g} H^2\,.
\end{equation}
Note that $ \frac{\partial F}{\partial \lambda}$ and $ \frac{\partial F}{\partial \eta}$ are related respectively to the integrated one-point functions
of the renormalized $\phi^4$ and $\mathcal{R}\phi^2$ operators. In fact, an unintegrated version of this equation (up to equations of motion and
total derivative terms, and including the Weyl square term as well) also holds as an operator equation on a general manifold. \footnote{See for instance eq. 10.1 of \cite{Hathrell:1981zb}.} In order
to have a scale invariant theory in curved space in $d=4-\epsilon$, we see therefore that we should set to zero not only the usual beta function for
$\lambda$, but also all curvature coupling beta functions\footnote{Of course, from a calculation of $F$ on $S^d$ we are not sensitive to $\beta_a$,
but we have included it here since it would play a role more generally on a not conformally flat space.}
\begin{equation}
\beta_{\lambda}=\beta_a=\beta_b=\beta_c=\beta_{\eta}=0\,.
\label{all-betavanish}
\end{equation}
Using the expressions for the beta functions (\ref{betalam}), (\ref{all-beta}),
these conditions can be solved in $d=4-\epsilon$ to give
\begin{equation}
\begin{aligned}
\label{all-star}
&\lambda=\lambda_*\,,\qquad a_*=-\frac{\hat\beta_a(\lambda_*)}{\epsilon}\,,\qquad
b_*=-\frac{\hat\beta_b(\lambda_*)}{\epsilon}\,,\\
&c_* = -\frac{\hat\beta_c(\lambda_*)+\eta_*\hat \beta_{\kappa}(\lambda_*)+\eta_*^2 \beta_{\Lambda}(\lambda_*)}{\epsilon}\,,\qquad
\eta_* = -\frac{\hat\beta_{\eta}(\lambda_*)}{\gamma_{\phi^2}(\lambda_*)},
\end{aligned}
\end{equation}
where $\lambda_*=\frac{8\pi^2}{N+8}\epsilon+\ldots$ is the value solving $\beta_{\lambda}(\lambda_*)=0$
which defines the Wilson-Fisher fixed point in flat space. More explicitly, using (\ref{pole-coeff}) and (\ref{hat-beta})
\begin{equation}
\begin{aligned}
\label{star-curv}
&a_* = -\frac{a_{01}}{\epsilon}-3 a_{21}\frac{\lambda_*^2}{\epsilon}+\mathcal{O}(\epsilon^2)\,,\qquad
b_* = -\frac{b_{01}}{\epsilon}-5 b_{41}\frac{\lambda_*^4}{\epsilon}-6 b_{51}\frac{\lambda_*^5}{\epsilon}+\mathcal{O}(\epsilon^5)\,,\\
&c_* = -6 c_{51}\frac{\lambda_*^5}{\epsilon}+\mathcal{O}(\epsilon^5)\,,\qquad \eta_* =-\frac{4\eta_{41}\lambda_*^4}{\gamma_{\phi^2}(\lambda_*)}+\mathcal{O}(\epsilon^4)\,.
\end{aligned}
\end{equation}
Tuning all couplings to their fixed point values, we then get the sphere free energy at the IR fixed point
\begin{equation}
F_{\textrm{IR}}(d)=F(\lambda_*, a_*, b_*, c_*, \eta_*,\mu R)\,.
\end{equation}
We emphasize that the resulting $F_{\textrm{IR}}(d)$ is just a function of $d=4-\epsilon$, the radius and $\mu$ dependence drop out as a consequence of
setting all beta functions to zero. We will verify this explicitly by our perturbative calculation below.

Eq.~(\ref{all-beta}) is our definition of the Wilson-Fisher fixed point on a curved space. Note that it is important that we work
in $d=4-\epsilon$ with a non-zero $\epsilon$. If we take the $\epsilon \rightarrow 0$
limit first, then the curvature beta functions become independent of $a,b,c$ and it is not possible to set them all to zero. Indeed,
if we take $d\rightarrow 4$ first, we get (since $\lambda_*=0$ in $d=4$):
\begin{equation}
\int_{S^4} d^4 x \sqrt{g} \langle T^{\mu}_{\mu} \rangle =  b_{01}\int_{S^4} d^4 x \sqrt{g} E = -\frac{N}{90}\,,
\end{equation}
where we used $\int_{S^4} d^4 x \sqrt{g} E=64\pi^2$. This is the expected conformal $a$-anomaly of $N$ free massless scalars.\footnote{On a
general manifold the term proportional to the square of the Weyl tensor $a_{01}\int_{S^4} d^4 x \sqrt{g} W^2$ will of course also appear on the right
hand side.} A related fact is that if we dial to the fixed point (\ref{all-beta}) in $d=4-\epsilon$,
and then take $\epsilon \rightarrow 0$, the free energy diverges: it has a simple pole whose residue is related to the $d=4$ anomaly.
It follows that the function $\tilde F_{\textrm{IR}}(d) = -\sin(\frac{\pi d}{2})F_{\textrm{IR}}(d)$ has a smooth limit as $d\rightarrow 4$
proportional to the $a$-anomaly. Indeed, since the first term in (\ref{F-split}) is finite as $\epsilon \rightarrow 0$, we have
\begin{eqnarray}
\lim_{d\rightarrow 4} \tilde F_{\textrm{IR}}(d) &=& -\frac{\pi}{2}\hat\beta_b(\lambda_*=0) \int_{S^4} d^4 x \sqrt{g} E\cr
&=& \frac{\pi}{2}\frac{N}{360(4\pi)^2}\int_{S^4} d^4x \sqrt{g}E = -\frac{\pi}{2}\int_{S^4} d^4 x \sqrt{g} \langle T^{\mu}_{\mu} \rangle\,.
\end{eqnarray}
While here we are considering the scalar field theory which has a trivial free fixed point in $d=4$, the fact that
$\lim_{d\rightarrow 4} \tilde F_{\textrm{IR}}(d)$ reproduces the 4d $a$-anomaly coefficient is general and should apply also to cases
where there is a non-trivial fixed point in $d=4$.

\section{Perturbative calculation of $F$ for the $O(N)$ model in $d=4-\epsilon$}
\label{F-pert}
\begin{figure}[h!]
                \centering
                \includegraphics[width=15cm]{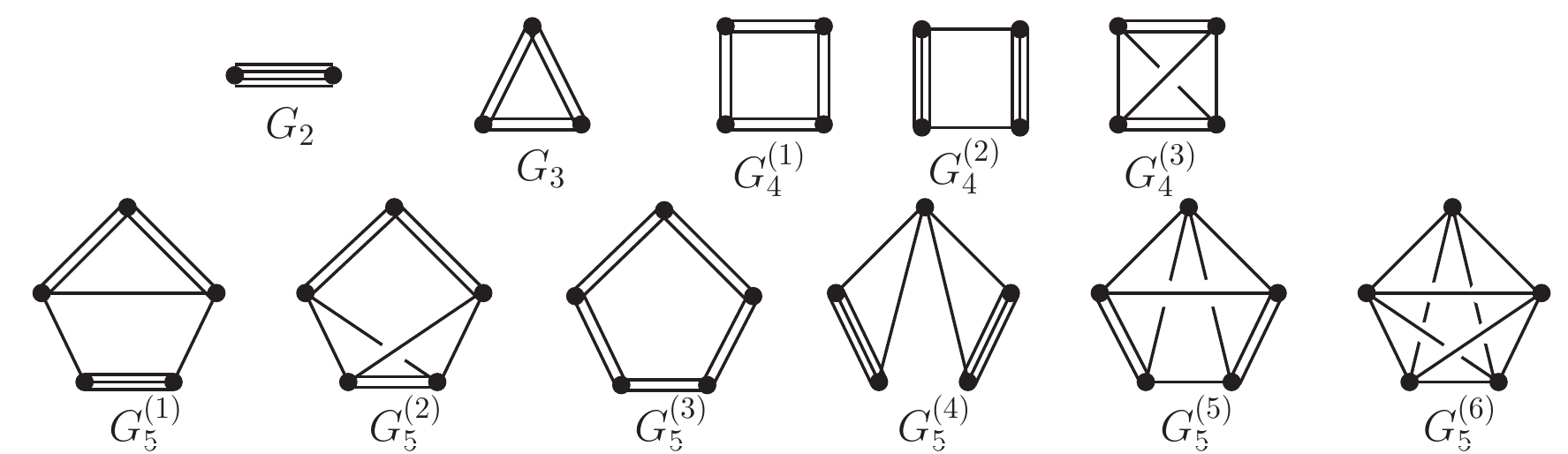}
               \caption{Diagrams contributing to the free energy in $\phi^{4}$-theory up to fifth order. Each line represents the sphere propogator $\langle \phi(x)\phi(y)\rangle = C_{\phi}/s(x,y)^{d-2}$. Symmetry factors are not included in these graphs.  }
\label{diagsPhi4}
\end{figure}

In this section we perform a calculation of $F$ at the IR fixed point of the $O(N)$ model in $d=4-\epsilon$ to order $\epsilon^4$, i.e.
we determine $\tilde F$ to order $\epsilon^5$ (the case of most general quartic coupling is worked out in Appendix \ref{gen-coup}).
Because of the structure of (\ref{star-curv}), this requires knowing the beta functions for
the curvature terms to order $\lambda^5$.  Let us first argue that to this order we can neglect the effect
of the renormalization of the conformal coupling parameter $\eta$. Note that once $F$ is expressed in terms
of renormalized parameters, it is finite for any value of $\eta$. Since the curvature counterterms $L_b, L_c$ in eq.~(\ref{curv-ren})
are independent of $\eta$, we can fix them by working at $\eta=0$. This means that we can set the bare parameter to $\eta_0 = L_{\eta}Z_2^{-1} =
\eta_{41}\frac{\lambda^4}{\epsilon}+\mathcal{O}(\lambda^5)$. The leading effect of this parameter is a divergence in $F$ of order $\frac{\lambda^6}{\epsilon}$,
coming from a two point function diagram ($G_2$ in figure \ref{diagsPhi4}) with a $\eta_0$ insertion in one of the propagators (the $0$-point and $1$-point
functions contribute to higher orders, respectively order $\frac{\lambda^8}{\epsilon}$ and $\frac{\lambda^8}{\epsilon^2}$). Hence, it does not
affect the calculation of $L_b$ and $L_c$ to the order $\lambda^5$ that we consider below. After the $\eta$-independent part of the counterterms
is fixed, we can reintroduce the $\eta$ dependence perturbatively. Recall that at the fixed point we set $\eta=\eta_*=\mathcal{O}(\epsilon^3)$,
see eq.~(\ref{star-curv}). The leading contribution to $F$ due to the $\eta_*$ is then a finite term of order $\eta_* \lambda_*^2\sim \mathcal{O}(\epsilon^5)$, coming from the two
point function with a ``mass" insertion on one propagator. In addition, the value of $c_*$ at the fixed point is shifted by the $\eta$ dependent terms
in (\ref{all-star}) only starting at order $\epsilon^6$. We conclude that, to the order we will be working on, we can ignore renormalization of the
conformal coupling parameter $\eta$ and work with the action
\begin{align}
S = \int d^{d}x \sqrt{g}\left(\frac{1}{2}\big((\partial_{\mu}\phi^{i}_{0})^{2}+\frac{d-2}{4(d-1)}\mathcal{R}(\phi_{0}^{i})^{2}\big)+\frac{\lambda_{0}}{4}(\phi_{0}^{i}\phi_{0}^{i})^{2}+b_{0}E +c_{0}H^{2} \right).
\end{align}

Introducing the integrated $n$-point functions of the free theory
\begin{align}
G_{n} = \int \prod_{i=1}^{n}d^{d}x_{i} \sqrt{g_{x_{i}}}\langle \phi_{0}^{4}(x_{1})...\phi_{0}^{4}(x_{n})\rangle^{\textrm{conn}}_{0}\,,
\end{align}
where $\phi_{0}^{4}=(\phi_{0}^{i}\phi_{0}^{i})^{2}$, the free energy up to fifth order is given by
\begin{align}
F-F_{\textrm{free}} = -\frac{\lambda_{0}^{2} }{2!\cdot 4^{2}}G_{2}+ \frac{\lambda_{0}^{3}}{3!\cdot 4^{3}} G_{3}- \frac{\lambda_{0}^{4}}{4!\cdot 4^{4}} G_{4}+ \frac{\lambda_{0}^{5}}{5!\cdot 4^{5}} G_{5}+\int d^{d}x\sqrt{g} (b_{0}E+c_{0}H^{2})\,, \label{freeEPhi4}
\end{align}
where $F_{\textrm{free}}$ is the contribution of the free field determinant \cite{Giombi:2014xxa}
\begin{equation}
F_{\textrm{free}}=N\log\det\left(-\nabla^2+\frac{d-2}{4(d-1)}\mathcal{R}\right)
=-N\int_0^1 du\, u\sin\pi u\, \frac{\Gamma\left(\frac{d}{2}+u\right)\Gamma\left(\frac{d}{2}-u\right)}{\sin(\frac{\pi d}{2})\Gamma\left(1+d\right)}
=\frac{N}{90\epsilon}+\mathcal{O}(\epsilon^0)\,.
\label{Ffree-div}
\end{equation}
The divergence of this determinant fixes $b_{01}=-\frac{N}{360(4\pi)^2}$ as given in (\ref{pole-coeff}). One can obtain the $\epsilon$-expansion of
$F_{\textrm{free}}$ to any desired order.

Using the two-point function
\begin{align}
\langle \phi^{i}(x) \phi^{j}(y)\rangle_{0} = \frac{C_{\phi}\delta^{ij}}{s(x,y)^{d-2}}\,,\qquad C_{\phi}=\frac{\Gamma(\frac{d}{2}-1)}{4\pi^{d/2}}\,,
\end{align}
where $s(x,y)$ is the chordal distance we find
\begin{align}
G_{2} &= 8N(N+2)C_{\phi}^{4}I_{2}(2(d-2))\,, \notag\\
G_{3}&= 64N(N+8)(N+2) C_{\phi}^{6}I_{3}(2d-4)\,, \notag\\
G_{4} &= 32(4!) N(N+2)\big((N^{2}+6N+20)G_{4}^{(1)}+8(N+2)G_{4}^{(2)} +4(5N+22)G_{4}^{(3)}\big)\,,\\
G_{5}&=2^{12}\cdot 3N(N+2)\Big(20 (N+2) (N+8) G_{5}^{(1)}+10  (3 N^{2}+22N+56)G_{5}^{(2)}\notag\\
&+ (N^3+8 N^2+24 N+48)G_{5}^{(3)}+10  (N+2)^2G_{5}^{(4)}+20 \left(N^2+20 N+60\right)G_{5}^{(5)}+8  (5 N+22)G_{5}^{(6)}\Big)\,,\notag
\label{phi4Combinatorics}
\end{align}
where all diagrams are represented in Figure \ref{diagsPhi4}.

\noindent The integrals $I_{2}(\Delta)$ and $I_{3}(\Delta)$ are given in Appendix \ref{Integ-App}, also using the method described in this appendix we find for the four point functions
\begin{align}
&G_{4}^{(k)} =\begin{cases}
\frac{4}{3 (4 \pi )^8 \epsilon ^2}+\frac{24 \big(\gamma+\log (4 \pi  R^2) \big)+73}{9 (4 \pi )^8 \epsilon }+\frac{5 \big(24 (\gamma+ \log (4 \pi  R^2) )+73\big)^2+360 \pi ^2+6713}{1080 (4 \pi )^8}+\mathcal{O}(\epsilon)\,, \\
-\frac{1}{20 (4 \pi )^8 \epsilon }-\frac{4 \big(\gamma+\log (4 \pi  R^2) \big)+15}{40 (4 \pi )^8}+\mathcal{O}(\epsilon)\,,\\
  \frac{2}{3 (4 \pi )^8 \epsilon ^2}+\frac{120 \big(\gamma +\log (4 \pi  R^2)\big)+383}{90 (4 \pi )^8 \epsilon }+\frac{\big(120 (\gamma+\log (4 \pi  R^2) )+383\big)^2+1800 \pi ^2+35041}{10800 (4 \pi )^8}+\mathcal{O}(\epsilon)\,,
\end{cases}
\end{align}
and  for the five point functions we get
\begin{align}
&G_{5}^{(k)} =\begin{cases}
-\frac{5}{3^3 (4 \pi )^{10} \epsilon ^2}-\frac{5 \big(30 (\gamma+\log (4 \pi  R^2) )+109\big)}{2^2 3^4 (4 \pi )^{10} \epsilon }+\mathcal{O}(\epsilon^{0})\,, \\
\frac{16}{9 (4 \pi )^{10} \epsilon ^3}+\frac{4 \big(10 (\gamma+\log (4 \pi  R^2) )+31\big)}{9 (4 \pi )^{10} \epsilon ^2}+\frac{3 \big(10 (\gamma+\log (4 \pi  R^2) )+31\big)^2+30 \pi ^2+641}{2\cdot 3^3 (4 \pi )^{10} \epsilon }+\mathcal{O}(\epsilon^{0})\,, \\
\frac{40}{9 (4 \pi )^{10} \epsilon ^3}+\frac{20 \big(15 (\gamma+\log (4 \pi  R^2) )+44\big)}{3^3 (4 \pi )^{10} \epsilon ^2}+\frac{10 \big(15 (\gamma+\log (4 \pi  R^2) )+44\big)^2+225 \pi ^2+4612}{2\cdot 3^4 (4 \pi )^{10} \epsilon }+\mathcal{O}(\epsilon^{0})\,,\\
\frac{2}{27 (4 \pi )^{10}}+\mathcal{O}(\epsilon)\,,\\
\frac{8}{9 (4 \pi )^{10} \epsilon ^3}+ \frac{4 \big(15 (\gamma+\log (4 \pi  R^2) )+49\big)}{3^3 (4 \pi )^{10} \epsilon ^2}+\frac{2 \big(15 (\gamma+\log (4 \pi  R^2) )+49\big)^2+45 \pi ^2+1014}{2\cdot 3^4 (4 \pi )^{10} \epsilon }+\mathcal{O}(\epsilon^{0})\,,\\
\frac{8 \zeta (3)}{9 (4 \pi )^{10} \epsilon }+\mathcal{O}(\epsilon^{0})\,.
\end{cases}
\end{align}
Combining all results in (\ref{freeEPhi4}), and expressing $\lambda_0, b_0, c_0$ in terms of renormalized couplings using (\ref{lam-bare}) and (\ref{curv-ren})-(\ref{curv-ren2})-(\ref{pole-coeff}), we find that all poles cancel provided we set the unknown $b_{51}$ coefficient in the
Euler term to
\begin{align}
b_{51}=\frac{N(N+2) \left(3(24+7N)+4(5 N+22) \zeta (3)\right)}{90 (4 \pi )^{12}}\,.
\label{b51-Euler}
\end{align}
This allows to find the Euler beta function $\beta_b$ to order $\lambda^5$, given in eq.~(\ref{betaEuler}), and hence determine the fixed point value $b_*$ to order $\epsilon^4$
by (\ref{star-curv}). Using the fixed point coupling
\begin{equation}
\begin{aligned}
&\lambda_* = \frac{8\pi^2}{N+8}\epsilon+\frac{24(3N+14)\pi^2}{(N+8)^3} \epsilon^2\\
&~~~~~~~-\frac{\pi ^2 \left(33 N^3-110N^2+96 (N+8) (5 N+22) \zeta (3)-1760N-4544\right)}{(N+8)^5}\epsilon^3+\ldots
\label{lamstar}
\end{aligned}
\end{equation}
In this equation $N\tilde{F}_s(4-\eps) = -\sin(\frac{\pi d}{2})F_{\rm free}$. Note that the dependence on $\mu R$ has dropped out,
consistently with the conformal invariance. As a partial check of this result, we note that at large $N$ it yields
\begin{equation}
\tilde{F}_{\textrm{IR}}-N \tilde{F}_s(4-\eps)=\left(-\frac{\pi}{576}\epsilon^3-\frac{13\pi}{6912}\epsilon^4+
\left(\frac{\pi^3}{13824}-\frac{647\pi}{414720}\right)\epsilon^5+\mathcal{O}(\eps^6)\right)+\mathcal{O}\left(\frac{1}{N}\right),
\end{equation}
which precisely agrees with the result derived in \cite{Giombi:2014xxa} using the Hubbard-Stratonovich transformation; it follows from (\ref{dtF}) with $\Delta=d-2$.

We can also obtain a result for $F$ in $d=4$ at a generic value of the renormalized coupling constant $\lambda$. Taking the limit $\epsilon \rightarrow 0$
of (\ref{freeEPhi4}) (after expressing all bare couplings in terms of renormalized ones), we find
\begin{align}
F^{d=4}-F^{d=4}_{\textrm{free}}&=-\frac{ N (N+2)\lambda^2}{72 (4 \pi )^4}+\frac{ N (N+2) (N+8) \big(5+2 (\gamma+\log (4 \pi  \mu ^2 R^2) )\big)\lambda^{3}}{72 (4 \pi )^6} -\frac{ N (N+2)\lambda^4}{1440 (4 \pi )^8}  \notag\\
& \times \Big(15 (N+8)^2 \big(5+2 (\gamma+\log (4 \pi  \mu ^2 R^2) )\big)^2+90 (3 N+14) \big(5+2 (\gamma +\log (4 \pi  \mu ^2 R^2))\big)\notag\\
&\qquad+63 N^2+1464 N+6272\Big)+ \frac{64}{3} \pi ^2 (3 b+2 c)\,.
 \end{align}
For the integrated trace of the energy-momentum tensor due to interactions we get
\begin{align}
&-\int_{S^4} d^{4}x \sqrt{g} \langle T_{\mu}^{\mu}\rangle = R \frac{\partial}{\partial R}(F^{d=4}-F^{d=4}_{\textrm{free}})\notag\\
&\qquad=\frac{ N (N+2) (N+8)\lambda^3}{18 (4 \pi )^6}-\frac{ N (N+2) \big(2 (N+8)^2 (\gamma+\log (4 \pi  \mu ^2 R^2) )+(5 N+89) N+362\big)\lambda^4}{12 (4 \pi )^8}\,.
 \end{align}
As a consistency check of this result, we note that it satisfies the Callan-Symanzik equation (\ref{CS}) to order $\lambda^4$ (without the $\beta_{\eta}$ term
that we have not included here).

\subsection{Pad\'e approximation of $\tilde{F}$}
\label{Pade-section}

For any quantity $f(d)$ known in the $\epsilon=4-d$ expansion up to a given order, we can construct a Pad\'e approximant
\begin{equation}
\textrm{Pad\'e}_{[m,n]}(\epsilon) = \frac{A_0 + A_1 \eps + A_2 \eps^2 + \ldots + A_m \eps^m}{1 + B_1 \eps + B_2 \eps^2 + \ldots + B_n \eps^n}\,,
\label{Pade}
\end{equation}
where the coefficients $A_i, B_i$ are fixed by requiring that the expansion at small $\eps$ agrees with the known terms in $f(4-\epsilon)$. If a quantity
is known in the $\eps$-expansion to order $k_0$, one can construct Pad\'e approximants of total order $m+n=k_0$.

An improved version of the Pad\'e approximant may be obtained if one knows a few terms of the $\eps$-expansion of the quantity $f(d)$
near two values of dimension:
\begin{align}
f(d_1+\eps)=& a_0 + a_1 \eps + a_2 \eps^2 + \ldots, \\
f(d_2-\eps)=&b_0+b_1\eps+b_2\eps^2 +\ldots\,.
\end{align}
Using the known expansions near the two boundaries, one can fix more coefficients in (\ref{Pade}) and hence use Pad\'e approximants of higher order, which should provide a better resummation of the quantity $f(d)$.

\begin{figure}[h!]
\centering
\includegraphics[width=10cm]{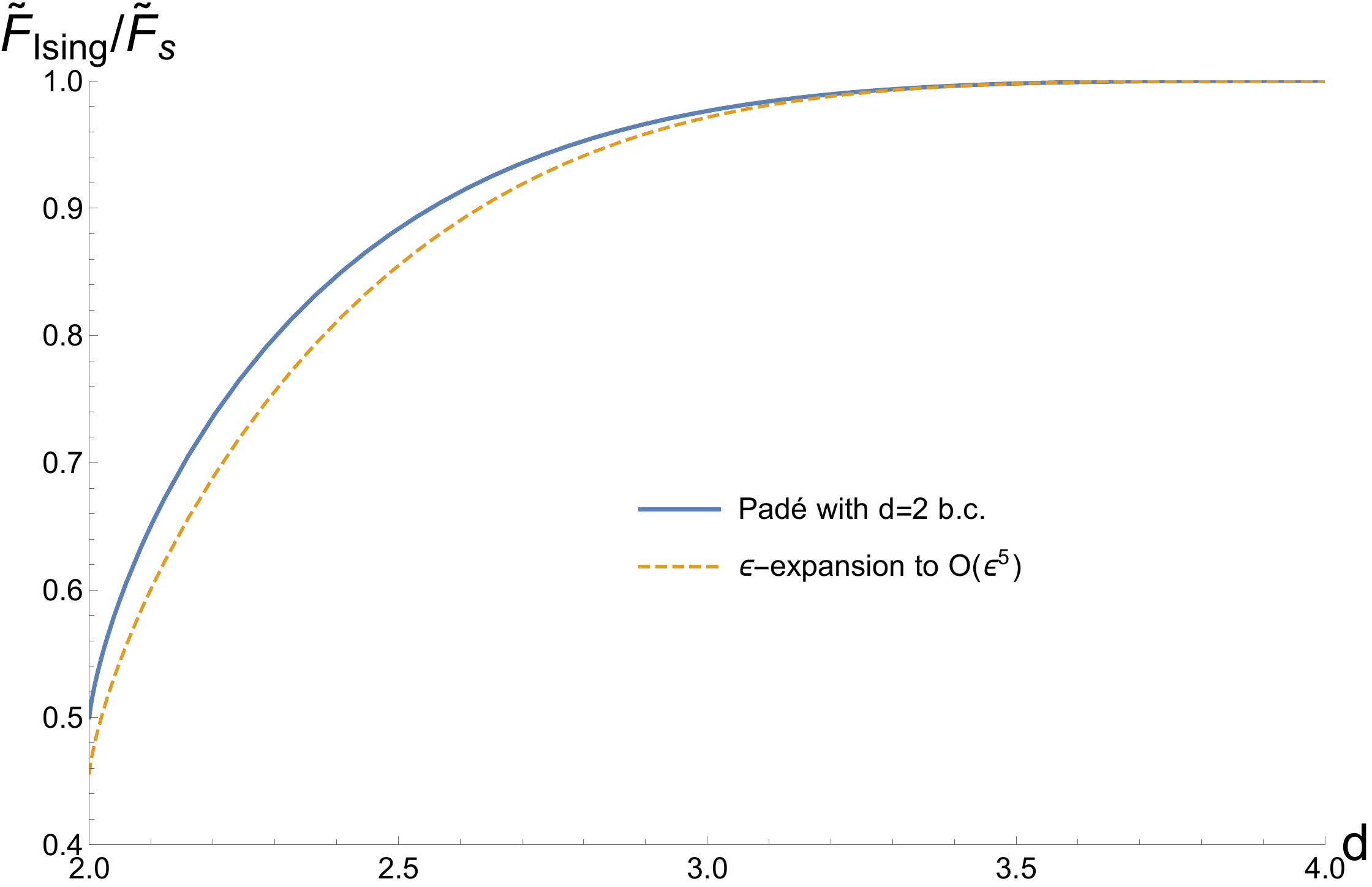}
\caption{$\tilde{F}/\tilde{F}_s$ for Ising model in $2\le d\le 4$. The solid line is the result obtained by averaging Pad\'e approximants
$\textrm{Pad\'e}_{[m,n]}$ with $m+n=6$ and fixed $d=2$ boundary condition.
The dashed line corresponds to the $\epsilon$ expansion result (\ref{F-Is-ep5}) without resummation.}
\label{Pade_Ising}
\end{figure}

For the Ising model ($N=1$) in $2\le d\le 4$, we know the expansion of $\tilde F$ in $d=4-\eps$ to order $\eps^5$, eq.~(\ref{tFON}).
In addition,
we know that for $d=2$ the IR fixed point of the quartic scalar field theory should coincide with the 2d Ising CFT, which is known to have central charge $c=1/2$.
This corresponds to the value $\tilde{F}_{\textrm{Ising}}(d=2)=\pi/12$. In other words, we have:
\begin{equation}
\tilde{F}_{\textrm{Ising}}(d) =
\begin{cases} \frac{\pi}{12} &\mbox{at }\quad d=2\,, \\
\tilde{F}_s(4-\eps)+\tilde{F}_{\textrm{int}} &\mbox{at }\quad d=4-\eps\,,
\end{cases}
\end{equation}
where
\begin{equation}
\label{F-Is-ep5}
{\tilde F}_s(4-\epsilon)+\tilde F_{\rm int}=\frac{\pi}{180} + 0.0205991 \eps + 0.0136429 \eps^2 + 0.00670643 \eps^3 + 0.00264883 \eps^4 + 0.000927589 \eps^5
+{\cal O}(\eps^6)\,.
\end{equation}
Imposing the $d=2$ value as boundary condition then allows us to use $\textrm{Pad\'e}_{[m,n]}$ approximants with total order $m+n=6$.
All approximants $\textrm{Pad\'e}_{[1,5]}$, $\textrm{Pad\'e}_{[2,4]}$, $\textrm{Pad\'e}_{[3,3]}$, $\textrm{Pad\'e}_{[4,2]}$and $\textrm{Pad\'e}_{[5,1]}$
appear to give very close results in the whole range $2\le d\le 4$.
In $d=3$, we get
\begin{equation}
\label{IsingPades}
\frac{\tilde{F}_{\textrm{Ising}}}{\tilde{F}_s} =
\begin{cases}
0.977973 &\quad \textrm{with}\quad \textrm{Pad\'e}_{[1,5]}\,, \\
0.97383 & \quad \textrm{with}\quad \textrm{Pad\'e}_{[2,4]}\,, \\
0.977106 &\quad \textrm{with}\quad \textrm{Pad\'e}_{[3,3]}\,, \\
0.977248 & \quad \textrm{with}\quad \textrm{Pad\'e}_{[4,2]}\,, \\
0.975379 &\quad \textrm{with}\quad \textrm{Pad\'e}_{[5,1]}\,.
\end{cases}
\end{equation}
Notice that all these values are lower than one, consistently with the $F$-theorem in $d=3$.
Given that these approximants are very close to each other, it seems reasonable to average over them. In this way, we obtain the estimate in $d=3$:
\begin{equation}
\label{Padeaverage}
\frac{\tilde{F}_{\textrm{Ising}}}{\tilde{F}_s} \approx 0.976\,.
\end{equation}
Using instead Pad\'e approximants of total order $m+n=5$, which are insensitive to the order $\eps^5$ in $\tilde F$ in $d=4-\epsilon$, we find
a very close estimate $\tilde{F}_{\textrm{Ising}}/ \tilde{F}_s \approx 0.974$, which indicates that the Pad\'e resummation appears to
be reliable.\footnote{Using the unresummed $\eps$-expansion (\ref{F-Is-ep5}) and setting $\eps=1$,
one obtains instead $\tilde{F}_{\textrm{Ising}}/ \tilde{F}_s \approx 0.957$ to order $\eps^4$ \cite{Giombi:2014xxa},
and $\tilde{F}_{\textrm{Ising}}/ \tilde{F}_s \approx 0.971$ to order $\eps^5$.}
In Figure \ref{Pade_Ising}, we plot $\tilde F_{\rm Ising}$ normalized by the free scalar result $\tilde{F}_s$, comparing the result of the
two-sided Pad\'e approximants and the unresummed $\epsilon$-expansion.


In the general $O(N)$ case for $N > 2$, the IR fixed point of the quartic theory is known to have an equivalent description
as the UV fixed point of the non-linear $O(N)$ sigma model, which is weakly coupled in $d=2+\epsilon$. For $d=2$, one has a free theory of $N-1$ massless scalar, which
has $c=N-1$, or $\tilde{F}=(N-1)\pi/6$. The same $d=2$ value applies to the case $N=2$, where one has a compact scalar with $c=1$ (in this
case, though, the behavior of the model in $d=2+\epsilon$ is not known).

For $N>2$, the expansion of $\tilde F_{\rm O(N)}$ in $d=2+\eps$ is expected to take the form
\begin{equation}
\tilde F_{O(N)}(2+\epsilon)=(N-1)\tilde F_s(2+\epsilon)+a_1 \eps+a_2 \eps^2+\ldots \,.
\end{equation}
where the coefficients $a_1, a_2,\ldots$ may in principle be fixed by doing a perturbative calculation on $S^{2+\epsilon}$.
This suggests that it may be more natural to apply the Pad\'e resummation procedure not to the full $\tilde F_{O(N)}=N\tilde F_s+\tilde F_{\rm int}$,
but to the quantity
\begin{equation}
f(d) \equiv \tilde F_{O(N)}-(N-1)\tilde F_s = \tilde F_s +\tilde F_{\rm int}\,,
\label{fd}
\end{equation}
since this is expected to have an expansion in integer powers of $\eps$ both near $d=4$ and near $d=2$ (at least for $N>2$).\footnote{Note
that the expansion of the free scalar answer $\tilde F_s$ by itself contains $\log(\eps)$ terms in $d=2+\eps$.} The exact value at $d=2$
is $f(2)=0$, and in $d=4-\eps$ we can read off the expansion up to order $\eps^5$ from (\ref{tFON}). Once the two-sided
Pad\'e approximant of $f(d)$ is obtained, we can reconstruct the full $\tilde F_{O(N)}$ by adding the contribution of $N-1$ free
scalars, using eq.~(\ref{tFfree}).

The $d=3$ estimates of $\tilde F_{O(N)}$ for a few values of $N$ obtained by using the $\textrm{Pad\'e}_{[4,2]}$ two-sided approximant on (\ref{fd})
are given in Table \ref{tableF}. For comparison, we also list the values obtained from ordinary Pad\'e approximants (without $d=2$ constraint) carried
out on the full $\tilde F_{O(N)}$.

As a test of our results, we note that in the large $N$ limit, our Pad\'e extrapolation with $d=2$ boundary condition yields, in $d=3$:
\begin{equation}
\tilde F_{O(N)} \approx 0.063807 N-0.015398+\frac{0.016974}{N}+{\cal O}(1/N^2) \,.
\label{FON-largeN}
\end{equation}
The term of order $N^0$ is within one percent of the exact value $-\zeta(3)/(8\pi^2)\approx -0.0152242$, which may be obtained from eq.~(\ref{dtF}) by setting $\Delta=1$, $d=3$. We have verified a similar good agreement with (\ref{dtF}) over the whole range $2\le d\le 4$. The term of order $1/N$ is not known exactly in $d=3$, and it would be interesting to compute it directly by large $N$ methods and compare with our estimate (\ref{FON-largeN}).

\begin{table}[h]
\centering
\begin{tabular}{ccccccc}
\hline
\multicolumn{1}{|c|}{N}         & \multicolumn{1}{c|}{1} & \multicolumn{1}{c|}{2} & \multicolumn{1}{c|}{3} & \multicolumn{1}{c|}{4}& \multicolumn{1}{c|}{5} & \multicolumn{1}{c|}{10} \\ \hline
\multicolumn{1}{|c|}{Pad\'e (b.c.)}      & \multicolumn{1}{c|}{0.9763} & \multicolumn{1}{c|}{0.9740} & \multicolumn{1}{c|}{0.9748} & \multicolumn{1}{c|}{0.9759} & \multicolumn{1}{c|}{0.9772}& \multicolumn{1}{c|}{0.9831}  \\ \hline
\multicolumn{1}{|c|}{Pad\'e (free)} & \multicolumn{1}{c|}{0.9779} & \multicolumn{1}{c|}{0.9765} & \multicolumn{1}{c|}{0.9762} & \multicolumn{1}{c|}{0.9766} & \multicolumn{1}{c|}{0.9773} & \multicolumn{1}{c|}{0.9820}  \\ \hline
\multicolumn{1}{l}{}            & \multicolumn{1}{l}{}   & \multicolumn{1}{l}{}   & \multicolumn{1}{l}{}   & \multicolumn{1}{l}{}   & \multicolumn{1}{l}{}  & \multicolumn{1}{l}{}
\end{tabular}
\caption{List of Pad\'e extrapolations for $\tilde{F}_{O(N)}/(N \tilde{F}_s)$ in $d=3$. The first line corresponds to the constrained Pad\'e approximants
with $d=2$ boundary condition, and the second line to ordinary Pad\'e approximants. For $N=1$ (b.c.) the entry is the average of several Pad\'e approximants, eq. (\ref{Padeaverage}). For $N\ge 2$, the
constrained Pad\'e is carried out on the quantity $f(d)$ defined in (\ref{fd}), as explained in the text. The ``free" Pad\'e approximant is carried out
on the full $\tilde F_{O(N)}$, eq.~(\ref{tFON}), using $\textrm{Pad\'e}_{[3,2]}$.}
\label{tableF}
\end{table}

We note that that the ratio $\tilde{F}_{O(N)}/(N \tilde{F}_s)$ in $d=3$, given in Table \ref{tableF}, has a minimum around $N=2$ using the constrained two-sided
Pad\'e, while it has a minimum close to $N=3$ using the unconstrained Pad\'e.\footnote{The latter result is closer to what was observed in \cite{Giombi:2014xxa}
using the unresummed $\epsilon$ expansion.} We observe that a similar behavior, with a mimimum close to $N=2$, was found in the conformal bootstrap approach \cite{Kos:2013tga} for the ratio $c^{3d~O(N)}_T/(N c_{T}^s)$. The estimates for $F$ given in Table \ref{tableF} imply in particular that, in $d=3$
\begin{equation}
N F_{\rm Ising} >F_{O(N)} \qquad {\rm for}~~N \lesssim 4\,.
\label{N-Ising-to-ON-Pa}
\end{equation}
As a consequence of the $d=3$ $F$-theorem, this suggests that for $N \gtrsim 5$
one cannot flow from $N$ copies of the 3d Ising CFT to the $O(N)$ invariant fixed point, while such a flow is possible for $N \lesssim 4$.
In section \ref{anisotropy} we will see that a similar behavior can be deduced in $d=4-\eps$ by studying an $O(N)$ breaking deformation of the quartic scalar field theory.

As an additional check of our results, note that by adding a relevant operator
(the mass term with a negative $m^2$) it is possible to make the CFT flow from the $O(N)$ invariant fixed point to
the phase where the $O(N)$ symmetry is broken to $O(N-1)$. In this phase
there are $N-1$ Goldstone bosons, and
the sphere free energy far in the IR is
\begin{equation}
F_{\rm SB} = (N-1)F_s = (N-1)\left(\frac{\log 2}{8}-\frac{3\zeta(3)}{16\pi^2}\right)\,.
\end{equation}
The F-theorem then requires that
\begin{equation}
F_{O(N)} > F_{\rm SB}\,.
\end{equation}
 For large $N$, this inequality was shown to hold in \cite{Klebanov:2011gs}. Using our results obtained from the $\epsilon$ expansion
and Pad\'e extrapolation, we have verified that it in fact holds for all $N$.


\section{$\tilde F$ in Conformal Perturbation Theory}
\label{confpert}

Let us consider a conformal field theory $\textrm{CFT}_0$ in dimension $d$ perturbed by a weakly relevant primary scalar operator
\begin{equation}
S_g = S_{\textrm{CFT}_0}+g_b \int d^d x\, O(x)\,.
\label{pertCFT}
\end{equation}
Here $g_b$ is a bare coupling constant, and the bare operator $O$ has dimension $\Delta=d-\epsilon$ in the unperturbed CFT,
with $\epsilon \ll 1$. The operator $O$ has flat space two- and three-point functions given by (in the unperturbed theory)
\begin{equation}
\begin{aligned}
&\langle O(x)O(y) \rangle_0 = \frac{{\cal C}_2}{|x-y|^{2\Delta}}\,,\\
&\langle O(x)O(y)O(z)\rangle_0= \frac{{\cal C}_3}{|x-y|^{\Delta} |y-z|^{\Delta} |z-x|^{\Delta}}\,.
\label{2pt3pt}
\end{aligned}
\end{equation}
The renormalization of the coupling constant can be deduced from a perturbative calculation of the two-point function in flat space. We have
\begin{eqnarray}
\langle O(x)O(y)\rangle_g &=& \langle O(x)O(y)\rangle_0 - g_b \int d^d z \langle O(x)O(y)O(z)\rangle_0 +\mathcal{O}(g_b^2)\notag\\
&=& \frac{{\cal C}_2}{|x-y|^{2\Delta}}-g_b \frac{{\cal C}_3}{|x-y|^{\Delta}}\int d^d z \frac{1}{|y-z|^{\Delta} |z-x|^{\Delta}}+\mathcal{O}(g_b^2)\,.
\end{eqnarray}
We can use the integral
\begin{equation}
\int d^d z\frac{1}{|x-z|^{2\alpha} |y-z|^{2\beta}} = \frac{\pi^{\frac{d}{2}}}{|x-y|^{2(\alpha+\beta-\frac{d}{2})}} \frac{\Gamma\left(\frac{d}{2}-\alpha\right)\Gamma\left(\frac{d}{2}-\beta\right)
\Gamma\left(\alpha+\beta -\frac{d}{2}\right)}{\Gamma\left(\alpha\right)\Gamma\left(\beta\right)\Gamma\left(d-\alpha-\beta\right)}\,,
\label{II}
\end{equation}
so we find, after plugging $\Delta=d-\epsilon$,
\begin{equation}
\langle O(x)O(y)\rangle_g = \frac{{\cal C}_2}{|x-y|^{2(d-\epsilon)}}-g_b \frac{{\cal C}_3 |x-y|^{\epsilon}}{|x-y|^{2(d-\epsilon)}}
\pi^{\frac{d}{2}}\frac{\Gamma\left(\frac{\epsilon}{2}\right)^2\Gamma\left(\frac{d}{2}-\epsilon\right)}
{\Gamma\left(\frac{d-\epsilon}{2}\right)^2\Gamma\left(\epsilon\right)}+\mathcal{O}(g_b^2)\,.
\end{equation}
In the limit $\epsilon\rightarrow 0$, we get
\begin{equation}
\langle O(x)O(y)\rangle_g = \frac{{\cal C}_2}{|x-y|^{2(d-\epsilon)}} - g_b \frac{{\cal C}_3} {|x-y|^{2(d-\epsilon)}}
\frac{4\pi^{\frac{d}{2}}}{\Gamma\left(\frac{d}{2}\right)}\frac{|x-y|^{\epsilon}}{\epsilon}+\mathcal{O}(g_b^2)\,.
\label{OObare}
\end{equation}
We can introduce a dimensionless renormalized coupling $g$ by
\begin{eqnarray}
g_b = \mu^{\epsilon}\left(g+z_1\frac{g^2}{\epsilon}+\ldots\right),
\label{gren}
\end{eqnarray}
where $\mu$ is the renormalization scale. This corresponds to the renormalization of the local operator
\begin{equation}
O(x) = Z_O O_{\rm ren}(x)\,,\qquad Z_O^{-1} = \frac{\partial (\mu^{-\epsilon}g_b)}{\partial g} =
1+2 z_1 \frac{g}{\epsilon}+\mathcal{O}(g^2)\,.
\end{equation}
We can fix $z_1$ by requiring that the two-point function of the renormalized operator is finite in the limit
$\epsilon \rightarrow 0$. From (\ref{OObare}) we find
\begin{equation}
\langle O_{\rm ren}(x)O_{\rm ren}(y)\rangle_g=\frac{1}{|x-y|^{2(d-\epsilon)}}
\left[{\cal C}_2(1+4z_1\frac{g}{\epsilon})-g {\cal C}_3\frac{4\pi^{\frac{d}{2}}}{\Gamma\left(\frac{d}{2}\right)}\frac{(\mu|x-y|)^{\epsilon}}{\epsilon}
+\mathcal{O}(g^2)\right]\,.
\label{OOren}
\end{equation}
This gives
\begin{equation}
z_1 = \frac{{\cal C}_3}{{\cal C}_2} \frac{\pi^{\frac{d}{2}}}{\Gamma\left(\frac{d}{2}\right)}\,.
\label{z1}
\end{equation}
From (\ref{gren}) we then find the beta function\cite{Cardy:1988cwa,Klebanov:2011gs}
\begin{equation}
\beta(g) = -\epsilon g+ \frac{\pi^{\frac{d}{2}}}{\Gamma\left(\frac{d}{2}\right)}  \frac{{\cal C}_3}{{\cal C}_2} g^2+\mathcal{O}(g^3)\,.
\label{betaPert}
\end{equation}
Then there is a IR stable fixed point given by
\begin{equation}
g_* = \frac{\Gamma\left(\frac{d}{2}\right)}{\pi^{\frac{d}{2}}}\frac{{\cal C}_2}{{\cal C}_3}\epsilon+\mathcal{O}(\epsilon^2)\,.
\label{gstar}
\end{equation}
The dimension of the renormalized operator $O_{\rm ren}(x)$ at the fixed point is given by
\begin{equation}
\Delta_{\textrm{IR}} = d-\epsilon+\beta(g)\frac{\partial}{\partial g}\log Z_O|_{g=g_*} =
d+\beta'(g_*) = d+\epsilon + \mathcal{O}(\epsilon^2)\,.
\end{equation}
Note that the operator has become irrelevant at the IR fixed point. Setting $g=g_*$ in (\ref{OOren}), we find that the two
point function of the renormalized operator takes the form, to leading order
\begin{equation}
\langle O_{\rm ren}(x) O_{\rm ren}(y)\rangle_{\textrm{IR}}
= \mu^{2(d-\epsilon)}\frac{\left({\cal C}_2+\mathcal{O}(\epsilon^2)\right)}{(\mu |x-y|)^{2(d+\epsilon+\mathcal{O}(\epsilon^2))}}\,.
\label{OOIR}
\end{equation}
Note that the leading correction to the two-point function normalization is expected to be of order $\epsilon^2$.

\subsection{The sphere free-energy}
We now conformally map the CFT (\ref{pertCFT}) to the sphere $S^d$. The two and three point functions of the bare operator
are the same as in (\ref{2pt3pt}) with the replacements $|x-y|\rightarrow s(x,y)$, etc., where $s(x,y)$ is
the chordal distance
\begin{equation}
s(x,y) = \frac{2R|x-y|}{(1+x^2)^{1/2}(1+y^2)^{1/2}}\,.
\end{equation}
and the metric of the sphere is $ds^2= \frac{4R^2 dx^{\mu}dx^{\mu}}{(1+x^2)^2}$.

Working in perturbation theory, we can compute the change in the free energy in the perturbed CFT to be
\cite{Klebanov:2011gs}
\begin{equation}
\begin{aligned}
\delta F &= F-F_0= -\frac{g_b^2}{2} {\cal C}_2 I_2(d-\epsilon)+\frac{g_b^3}{6} {\cal C}_3 I_3(d-\epsilon)+\mathcal{O}(g_b^4)\\
&=-\frac{\mu^{2\epsilon}g^2}{2} {\cal C}_2 I_2(d-\epsilon)+\mu^{2\epsilon}g^3 {\cal C}_3
\left(\frac{1}{6}\mu^{\epsilon}I_3(d-\epsilon)
-\frac{\pi^{\frac{d}{2}}}{\epsilon\Gamma\left(\frac{d}{2}\right)}I_2(d-\epsilon)\right)+\mathcal{O}(g^4)\,,
\end{aligned}
\end{equation}
where in the second line we have used the relation (\ref{gren}),(\ref{z1}) between bare and renormalized coupling, and
we have used the two and three point integrals on the sphere $I_{2}(\Delta)$ and $I_{3}(\Delta)$, which are given in Appendix \ref{Integ-App}.
Expanding at small $\epsilon$ with fixed $d$, we find
\begin{equation}
I_2(d-\epsilon) = \frac{2^{1-d}\pi^{d+\frac{1}{2}}\Gamma\left(-\frac{d}{2}\right)}{\Gamma\left(\frac{1+d}{2}\right)}\epsilon+\mathcal{O}(\epsilon^2)\,,\qquad I_3(d-\epsilon) = \frac{8\pi^{\frac{3d}{2}}\Gamma\left(-\frac{d}{2}\right)}{\Gamma\left(d\right)}+\mathcal{O}(\epsilon)\,. \label{I2I3ep}
\end{equation}
Note that
\begin{equation}
\frac{I_3(d-\epsilon)}{I_2(d-\epsilon)}=\frac{8\pi^{\frac{d}{2}}}{\epsilon\Gamma\left(\frac{d}{2}\right)}+\mathcal{O}(\epsilon^0)\,.
\end{equation}
To leading order in $\epsilon$ and $g$ we then find, using (\ref{I2I3ep})
\begin{equation}
\delta F = \frac{2^{1-d}\pi^{d+\frac{1}{2}}\Gamma\left(-\frac{d}{2}\right)}{\Gamma\left(\frac{1+d}{2}\right)}
{\cal C}_2
\left[-\frac{1}{2}\epsilon g^2+\frac{1}{3}\frac{\pi^{\frac{d}{2}}}{\Gamma\left(\frac{d}{2}\right)}\frac{{\cal C}_3}{{\cal C}_2} g^3\right]\,.
\end{equation}
Finally, in terms of $\tilde F=-\sin( \frac{\pi d}{2})F$, after some simplification of the $d$-dependent prefactor
we obtain
\begin{equation}
\delta \tilde F = \frac{2{\cal C}_2\pi^{1+d}}{\Gamma\left(1+d\right)}
\left[-\frac{1}{2}\epsilon g^2+\frac{1}{3}\frac{\pi^{\frac{d}{2}}}{\Gamma\left(\frac{d}{2}\right)}\frac{{\cal C}_3}{{\cal C}_2} g^3\right]\,.
\end{equation}
Note that to this order we find
\begin{equation}
\frac{\partial \delta \tilde F}{\partial g} =\frac{2{\cal C}_2\pi^{1+d}}{\Gamma\left(1+d\right)}\beta(g)
\label{tF-der}
\end{equation}
and so $\tilde F$ is stationary at the fixed point. To evaluate the change in $\tilde F$ from UV to IR, we
can plug in the value of the fixed point coupling (\ref{gstar}), and we get
\begin{equation}
\delta \tilde F  = \tilde F_{\textrm{IR}}-\tilde F_{\textrm{UV}}=-\frac{\pi \Gamma\left(\frac{d}{2}\right)^2}{\Gamma\left(1+d\right)}
\frac{{\cal C}_2^{3}}{3 {\cal C}_3^2}\epsilon^3\,.
\label{deltaF}
\end{equation}
In odd $d$ this result agrees with \cite{Klebanov:2011gs},
and in even $d$ with the change in $a$-anomaly computed in \cite{Komargodski:2011xv}.
For a unitary CFT, ${\cal C}_2>0$ and ${\cal C}_3$ is real. So we find that for all $d$, $\tilde F_{\textrm{UV}}>\tilde F_{\textrm{IR}}$ to leading
order in conformal perturbation theory.
This gives a perturbative proof of the Generalized $F$-theorem in the framework of conformal perturbation theory.

Note that using (\ref{tF-der}) we may rewrite the leading order change in $\tilde F$ as
\begin{equation}
\delta \tilde F =\frac{2\pi^{1+d}}{\Gamma\left(1+d\right)}\int_0^{g_*} {\cal C}_2\,\beta(g) dg\,.
\label{intBeta}
\end{equation}
It is natural  to ask whether such a relation continues to hold at higher orders, provided we replace ${\cal C}_2$ by a
coupling dependent two-point function coefficient ${\cal C}_2(g)$.

\subsection{Several coupling constants}
\label{confseveral}

More generally, we can consider a perturbation by several primary operators
\begin{equation}
S_g = S_{\textrm{CFT}_0}+\sum_i g^i_b \int d^d x \,O_i(x) \,,
\end{equation}
where $g^i_b$ are bare coupling constants. For simplicity, we take the bare operators $O_i$ to
have the same dimension
\begin{equation}
\Delta=d-\epsilon\,,
\end{equation}
where as before we assume $\epsilon \ll 1$.

The two and three point functions in flat space take the form
\begin{equation}
\begin{aligned}
&\langle O_i(x)O_j(y) \rangle_0 = \frac{{\cal  C}_{2, ij}}{|x-y|^{2\Delta}}\,,\\
&\langle O_i(x)O_j(y)O_k(z)\rangle_0= \frac{{\cal C}_{ijk}}{|x-y|^{\Delta} |y-z|^{\Delta}
|z-x|^{\Delta}} \,.
\label{2pt3pt2}
\end{aligned}
\end{equation}
Here ${\cal C}_{2,ij}$ is a symmetric matrix (positive in unitary theories). We could choose a basis of operators
so that ${\cal C}_{2,ij}=\delta_{ij}$, but we will not do this here.

Let us consider the calculation of the two-point function in the perturbed CFT. We have
\begin{eqnarray}
\langle O_i(x)O_j(y)\rangle_g &=& \langle O_i(x)O_j(y)\rangle_0 - \sum_k g^k_b \int d^d z \langle O_i(x)O_j(y)O_k(z)\rangle_0 +\mathcal{O}(g_b^2) \notag \\
&=& \frac{{\cal  C}_{2, ij}}{|x-y|^{2\Delta}}-\sum_k g^k_b \frac{{\cal C}_{ijk}}{|x-y|^{\Delta}}
\int d^d z \frac{1}{|y-z|^{\Delta} |z-x|^{\Delta}}+\mathcal{O}(g_b^2)\,.
\end{eqnarray}
Using the integral (\ref{II}), and expanding at small $\epsilon$, we find
\begin{equation}
\langle O_i(x)O_j(y)\rangle_g  =
\frac{{\cal  C}_{2, ij}}{|x-y|^{2(d-\epsilon)}}
-\sum_k g^k_b \frac{{\cal C}_{ijk}}{|x-y|^{2(d-\epsilon)}}
\frac{4\pi^{\frac{d}{2}}|x-y|^{\epsilon}}{\Gamma\left(\frac{d}{2}\right)}\frac{1}{\epsilon}+\mathcal{O}(g_b^2)\,.
\label{OOMany}
\end{equation}
We can introduce the dimensionless renormalized couplings $g_i$ by
\begin{eqnarray}
g^i_b = \mu^{\epsilon}\left(g^i+\sum_{jk} z^i_{jk}\frac{g^j g^k}{\epsilon}+\ldots\right)\,.
\label{grenMany}
\end{eqnarray}
The corresponding local operators renormalization is
\begin{equation}\
\begin{aligned}
&O_i(x) = (Z_O)_{ij} O^{\rm ren}_j(x) \,,\\
& O^{\rm ren}_k(x) =\left(\delta^i_k+2z^i_{jk}g^j\right)O_i(x) \,.
\end{aligned}
\end{equation}
Plugging this into (\ref{OOMany}) and requiring that the renormalized two-point function has no poles as $\epsilon \rightarrow 0$, we get
the condition
\begin{equation}
2{\cal C}_{2,ik}z^k_{lm}g^m+2{\cal C}_{2,jk}z^k_{im}g^m
-\frac{4\pi^{\frac{d}{2}}}{\Gamma\left(\frac{d}{2}\right)}{\cal C}_{ijm}g^m=0\,.
\end{equation}
This is solved by
\begin{equation}
\begin{aligned}
z^i_{jk} = {\cal C}_2^{il}{\cal C}_{ljk}\frac{\pi^{\frac{d}{2}}}{\Gamma\left(\frac{d}{2}\right)}\,,
\end{aligned}
\end{equation}
where ${\cal C}_2^{il}$ is the inverse of the matrix ${\cal C}_{2,il}$.
The beta functions are then given by
\begin{equation}
\beta^i =\mu \frac{dg^i}{d\mu} =
-\epsilon g^i+\frac{\pi^{\frac{d}{2}}}{\Gamma\left(\frac{d}{2}\right)} {\cal C}_2^{il}{\cal C}_{ljk}g^j  g^k+\mathcal{O}(g^3)\,.
\label{betai}
\end{equation}

The computation of the sphere free energy proceeds as in the single coupling case described above. In terms
of the integrals $I_{2}$ and $I_{3}$ we have
\begin{equation}
\begin{aligned}
\delta F &= F-F_0= -\frac{g_b^i g_b^j}{2} {\cal C}_{2,ij} I_2(d-\epsilon)
+\frac{g_b^i g_b^j g_b^k}{6} {\cal C}_{ijk} I_3(d-\epsilon)+\mathcal{O}(g_b^4)\\
&=\frac{2^{1-d}\pi^{d+\frac{1}{2}}\Gamma\left(-\frac{d}{2}\right)}{\Gamma\left(\frac{1+d}{2}\right)}
\left[-\frac{1}{2}\epsilon g^i g^j {\cal C}_{2,ij}
+\frac{1}{3}\frac{\pi^{\frac{d}{2}}}{\Gamma\left(\frac{d}{2}\right)}{\cal C}_{ijk} g^i g^j g^k+\mathcal{O}(g^4)\right],
\end{aligned}
\end{equation}
where in the second line we have used the small $\epsilon$ expansion of $I_2$ and $I_3$, and we have plugged in
the relation between bare and renormalized couplings. In terms of $\tilde F$, we have
\begin{equation}
\delta \tilde F = \frac{2\pi^{1+d}}{\Gamma\left(1+d\right)}
\left[-\frac{1}{2}\epsilon g^i g^j {\cal C}_{2,ij}
+\frac{1}{3}\frac{\pi^{\frac{d}{2}}}{\Gamma\left(\frac{d}{2}\right)}{\cal C}_{ijk} g^i g^j g^k\right]\,.
\end{equation}
Analogously to the one-coupling case, we then find
\begin{equation}
\frac{\partial \delta \tilde F}{\partial g^i}=
\frac{2\pi^{1+d}}{\Gamma\left(1+d\right)} {\cal C}_{2,ij}\beta^j\,.
\label{Fvsbeta}
\end{equation}
To write the final result more explicitly, we note that the vanishing of the beta functions implies
\begin{equation}
0=-\epsilon g_*^i g_*^j {\cal C}_{2,ij}+\frac{\pi^{\frac{d}{2}}}{\Gamma\left(\frac{d}{2}\right)}{\cal C}_{ijk}g_*^i g_*^j g_*^k
\end{equation}
and so we find
\begin{equation}
\delta \tilde F = \tilde F_{\textrm{IR}}-\tilde F_{\textrm{UV}}  = -\frac{\pi^{1+d}}{3\Gamma\left(1+d\right)} {\cal C}_{2,ij}g^i_* g^j_*\epsilon
+\mathcal{O}(\epsilon^3)\,.
\label{dFij}
\end{equation}
In a unitary theory, the matrix ${\cal C}_{2,ij}$ is positive definite and the coupling constants are real, so we find as
expected $\tilde F_{\textrm{UV}}>\tilde F_{\textrm{IR}}$.

Note that (\ref{Fvsbeta}) implies that the leading order change in $\tilde F$ from UV to IR can be written as the line integral
\begin{equation}
\delta \tilde F = \tilde F_{\textrm{IR}}-\tilde F_{\textrm{UV}} =
\frac{2\pi^{1+d}}{\Gamma\left(1+d\right)}\int_0^{\vec{g}_*} {\cal C}_{2,ij}\beta^j dg^i \,,
\label{deltaFMany}
\end{equation}
where the integral is along a path connecting the origin in coupling space to the point $\vec{g}_*=(g^1_*,g^2_*,\ldots)$.
The integral is independent on the choice of path due to (\ref{Fvsbeta}), which implies
\begin{equation}
{\cal C}_{2,ij}\frac{\partial\beta^j}{\partial  g^k}={\cal C}_{2,kj}\frac{\partial\beta^j}{\partial  g^i}\,.
\end{equation}

\subsection{The $O(N)$ model in $4-\epsilon$ dimensions}

Let us consider the scalar field theory with $O(N)$ symmetric interaction in $d=4-\epsilon$
\begin{equation}
S = \int d^{d}x \left(\frac{1}{2}(\partial_{\mu}\phi_{i})^{2}+\frac{g}{4}(\phi_{i}\phi_{i})^{2}\right)\,.
\end{equation}
We can view the quartic interaction as a perturbation of the form (\ref{pertCFT}), where the unperturbed CFT is the free theory, and
the perturbing operator $O(x)=\frac{1}{4}(\phi_{i}\phi_{i})^{2}$ has dimension $\Delta=2d-4=d-\epsilon$ in the UV. Then we can
formally apply the results of the previous section to the present case. Note that in the above calculations
we have assumed a CFT defined in dimension $d$ with an operator of dimension $d-\epsilon$, and $\epsilon$ in principle unrelated to $d$.
Therefore, we are expanding in $\epsilon$ first with $d$ fixed, and at the end set $d=4-\epsilon$ (this corresponds to a different order
of limits compared to the approach in section \ref{F-pert}). In the free theory at dimension $d$, we have
\begin{eqnarray}
&&\langle O(x)O(y) \rangle_0 = \frac{1}{2}N(N+2) \left(\frac{\Gamma(d/2-1)}{4\pi^{d/2}}\right)^4\frac{1}{|x-y|^{2(d-\epsilon)}}\,,\\
&&\langle O(x)O(y)O(z) \rangle_0 =N(N+8)(N+2)\left(\frac{\Gamma(d/2-1)}{4\pi^{d/2}}\right)^6
\frac{1}{\left(|x-y||y-z||z-x|\right)^{d-\epsilon}}\,.\nonumber
\end{eqnarray}
From which we can read off, in the notation of the previous section
\begin{equation}
\begin{aligned}
&{\cal C}_2 = \frac{1}{2} N(N+2) \left(\frac{\Gamma(d/2-1)}{4\pi^{d/2}}\right)^4\,,\qquad
{\cal C}_3 = N(N+8)(N+2)\left(\frac{\Gamma(d/2-1)}{4\pi^{d/2}}\right)^6 \,,
\label{C2C3phi4}
\end{aligned}
\end{equation}
so that the conformal perturbation theory $\beta$-function reads
\begin{equation}
\beta = -\epsilon g + \frac{(N+8)\Gamma\left(\frac{d}{2}-1\right)^2}{8\pi^{\frac{d}{2}}\Gamma\left(\frac{d}{2}\right)}g^2+\mathcal{O}(g^3)\,.
\label{betaphi4}
\end{equation}
Note that setting $d=4$, this agrees with the well-known result for the beta function in the quartic scalar field theory computed
in the standard field theory approach, $\beta=-\epsilon g+\frac{N+8}{8\pi^2}g^2+\mathcal{O}(g^3)$.

To obtain the leading order change in $\tilde F$ on the sphere, we just need to apply the general result (\ref{deltaF}). Using (\ref{C2C3phi4}),
and setting $d=4-\epsilon$, we find
\begin{equation}
\delta \tilde F = -\frac{\pi}{576}\frac{N(N+2)}{(N+8)^2}\epsilon^3+\mathcal{O}(\epsilon^4)\,.
\end{equation}
This agrees precisely with the leading order term in (\ref{tFON}), first obtained in \cite{Giombi:2014xxa}.

In the conformal perturbation theory approach, it is not easy to deduce the next order correction
in the beta function (\ref{betaPert}), since it depends on the
4-point function of the perturbing operator. However, in the present case we know that for $d=4$ it should reduce
to the known beta function computed in the minimal subtraction scheme, eq.~(\ref{betalam}), namely we should have
\begin{equation}
\beta = -\epsilon g + \frac{(N+8)\Gamma\left(\frac{d}{2}-1\right)}{8\pi^{\frac{d}{2}}\Gamma\left(\frac{d}{2}\right)}g^2
-\left(\frac{3(3N+14)}{64\pi^4}+\mathcal{O}(\epsilon)\right)g^3+\mathcal{O}(g^4)\,.
\label{betag3}
\end{equation}
Assuming that the relation of the form (\ref{intBeta}) holds to higher orders, and using the fact that
${\cal C}_2(g)={\cal C}_2+\mathcal{O}(g^2)$ (see (\ref{OOIR})), we then find
\begin{equation}
\begin{aligned}
&\delta \tilde F = \frac{2\pi^{1+d}}{\Gamma\left(1+d\right)}\int_0^{g_*} {\cal C}_2
\left[-\epsilon g + \frac{(N+8)\Gamma\left(\frac{d}{2}-1\right)}{8\pi^{\frac{d}{2}}\Gamma\left(\frac{d}{2}\right)}g^2
-\left(\frac{3(3N+14)}{64\pi^4}\right)g^3\right]dg +\mathcal{O}(\epsilon^5)\\
&~~~~~~~= -\frac{\pi}{576}\frac{N(N+2)}{(N+8)^2}\epsilon^3
-\frac{\pi}{6912}\frac{N(N+2)(13N^2+370N+1588)}{(N+8)^4}\epsilon^4+\mathcal{O}(\epsilon^5)\,.
\label{ON-conf-pert}
\end{aligned}
\end{equation}
Remarkably, this agrees with the result (\ref{tFON}) obtained by the direct method based on the renormalization of the field
theory on curved space, provided one includes the effect of the Euler density counterterm which affects the
order $\epsilon^4$ in $\tilde F$.

\subsection{The $O(N)$ model with anisotropic perturbation}
\label{anisotropy}
\begin{figure}[h!]
\begin{center}
\includegraphics[width=10cm]{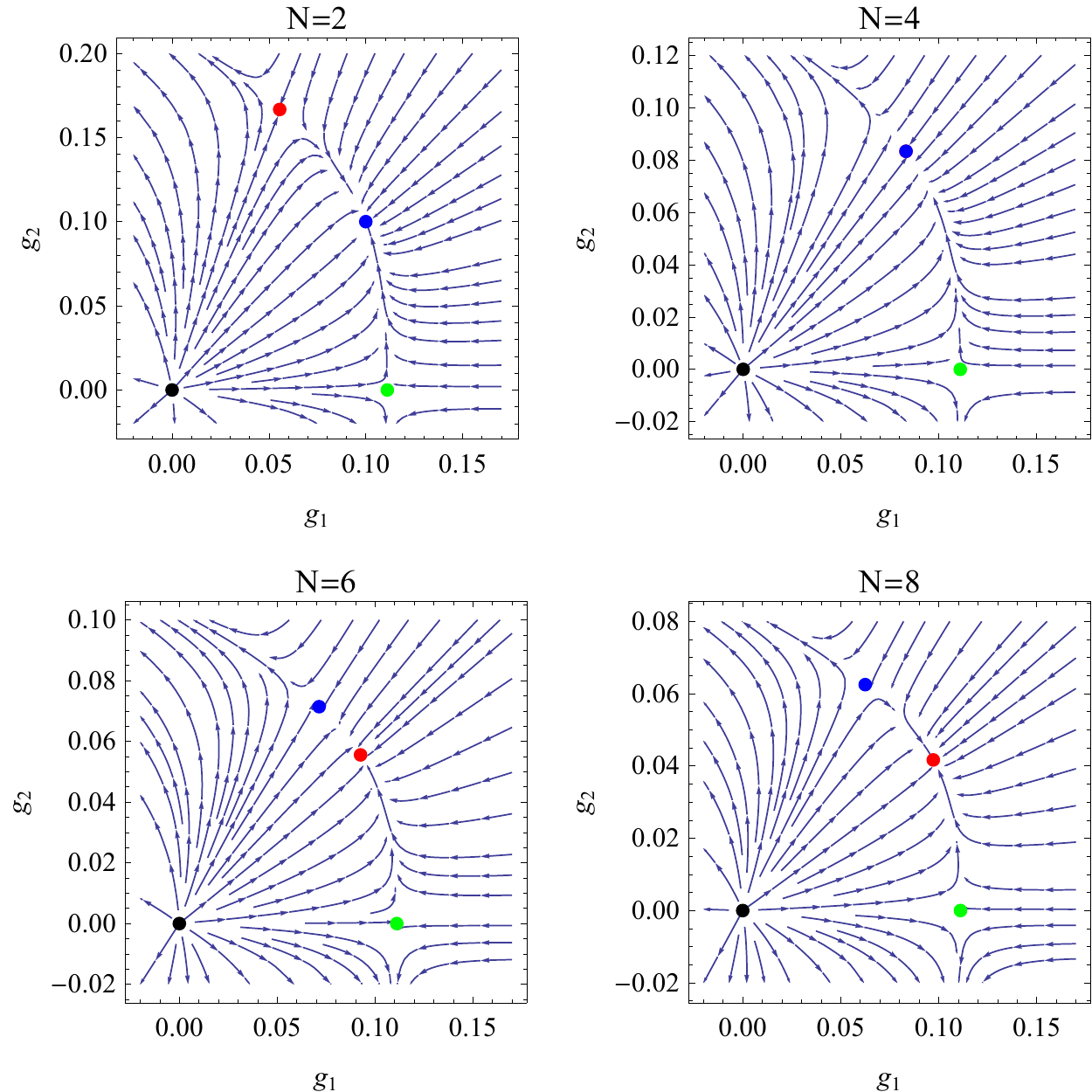}
\caption{Structure of the one-loop RG flow trajectories for $N=2,4,6,8$. The black dot indicates the free UV fixed point,
the blue one the $O(N)$ fixed point, the green one the $N$ decoupled Ising models, and the red one the anisotropic
fixed point. For $N<4$, the $O(N)$ fixed  point is IR stable, while for $N>4$ the cubic point is. For $N=4$,
the $O(N)$ and cubic fixed point coincide.}
\label{Flows}
\end{center}
\end{figure}

Let us consider the $O(N)$ scalar field theory in $d=4-\epsilon$ perturbed by an operator that breaks $O(N)$ to $S_N\times Z_2^N$, which are the symmetries of the $N$-dimensional hypercube. This is why it is sometimes called the model with ``cubic anisotropy" \cite{Kleinert:2001ax}:
\begin{equation}
S = \int d^{d}x \left(\frac{1}{2}(\partial_{\mu}\phi^{i})^{2}+\frac{\lambda_1}{4}(\phi^{i}\phi^{i})^{2}
+\frac{\lambda_2}{4}(\sum_i \phi_i^4)\right)\,.
\end{equation}
Both operators in the action, $(\phi^{i}\phi^{i})^{2}$ and $(\sum_i \phi_i^4)$,
have dimension $\Delta=2d-4=d-\epsilon$ in the free UV CFT, and so we can apply the results of
section \ref{confseveral} to this model. In this basis of operators the two-point function matrix ${\cal C}_{2,ij}$ is not diagonal. While
it is not necessary to do so, we find it convenient to change the basis to
\begin{equation}
S = \int d^{d}x \left(\frac{1}{2}(\partial_{\mu}\phi_{i})^{2}+\frac{g_1}{4}(\sum_i \phi_i^4)
+\frac{g_2}{2}\sum_{i<j}\phi_{i}^2\phi_{j}^2\right)\,.
\label{cubicS}
\end{equation}
In terms of the operators $O_1(x)=\frac{1}{4}(\sum_i \phi_i^4)$ and $O_2(x)=\frac{1}{2}\sum_{i<j}\phi_{i}^2\phi_{j}^2$,
the two-point functions in the free theory read
\begin{equation}
\begin{aligned}
&\langle O_i(x) O_j(y)\rangle = \frac{({\cal C}_2)_{ij}}{|x-y|^{2(d-\epsilon)}}\,,\\
&({\cal C}_2)_{11}=\frac{3}{2} N \left(\frac{\Gamma(d/2-1)}{4\pi^{d/2}}\right)^4\,,\qquad
({\cal C}_2)_{22}=\frac{N(N-1)}{2}\left(\frac{\Gamma(d/2-1)}{4\pi^{d/2}}\right)^4\,,\qquad ({\cal C}_2)_{12}=0\,.
\label{C2ZN}
\end{aligned}
\end{equation}
The one-loop beta functions computed with the standard dimensional
regularization and minimal subtraction scheme are known to be \cite{Kleinert:2001ax} (see \ref{cubic-App} for the higher order terms)
\begin{equation}
\begin{aligned}
&\beta_1 = -\eps g_{1} + \frac{9g_1^2+(N-1)g_2^2}{8\pi^2}\,, \\
&\beta_2 = -\eps g_{2} + \frac{(N+2)g_2^2+6g_1 g_2}{8\pi^2}\,.
\label{betaZN}
\end{aligned}
\end{equation}
Besides the free UV fixed point, there are three IR fixed points, given to the one-loop order by
\begin{equation}
\begin{aligned}
&g_1^* =g_2^*=\frac{8\pi^2}{N+8}\epsilon \,,\qquad ~~~~~~~~~~~~~~~~~~~~~\mbox{$O(N)$ theory}\\
&g_1^* =\frac{8\pi^2}{9}\epsilon \,,\qquad g_2^* = 0\,,\qquad ~~~~~~~~~~~~~~\mbox{$N$ decoupled Ising}\\
&g_1^* =\frac{8\pi^2(N-1)}{9N} \epsilon \,,\qquad g_2^* =\frac{8\pi^2}{3N}\epsilon \,,\qquad \mbox{Anisotropic fixed point}
\label{gsZN}
\end{aligned}
\end{equation}
The first one is the usual $O(N)$ invariant Wilson-Fisher fixed point, the second is the decoupled product of $N$ Ising fixed points,
and the last one is a new fixed point where the $O(N)$ symmetry is broken by the anisotropic term.
The eigenvalues $\lambda_{\pm}$ of the stability matrix $M_{ij}=\frac{\partial \beta_i}{\partial g_j}$, which are
related to the dimensions of the nearly marginal operators at the fixed point by $\Delta_{\pm}=d+\lambda_{\pm}$, are
given by
\begin{equation}
\begin{aligned}
O(N):&~~\lambda_+=\epsilon\,,\quad \lambda_-=\frac{(4-N)\epsilon}{N+8}\,,\\
N-{\rm Ising}:&~~\lambda_+=\epsilon\,,\quad \lambda_-=-\frac{\epsilon}{3}\,,\\
{\rm Anisotropic}:&~~\lambda_+=\epsilon\,,\quad \lambda_-=\frac{(N-4)\epsilon}{3N}\,.
\end{aligned}
\end{equation}
Thus we see that the $O(N)$ symmetric fixed point is IR stable for $N<N_c=4$, while the anisotropic one is IR stable for $N>N_c$.
The $N$-Ising fixed point always has one relevant operator which triggers a flow to the $O(N)$ symmetric fixed point for $N<N_c$
or to the anisotropic one for $N>N_c$. Plots of the RG flow trajectories for $N=2,4,6,8$ are given in  Fig. \ref{Flows}. The value
$N_c=4$ is the leading order one-loop result. Higher order corrections make $N_c$ decrease as we move away from $d=4$. In $d=3$,
it is expected that $N_c \simeq 3$ \cite{Kleinert:2001ax}.

We now compute the leading order correction to $\tilde F$ at each fixed point
and check that the structure of the RG flows described
above is consistent with the Generalized $F$-theorem. The beta functions (\ref{betaZN}) differ
from the conformal perturbation theory beta functions (\ref{betai}) by the $d$-dependent factors in the quadratic
terms. However, this difference does not affect the result for $\tilde F$ at leading order in $\epsilon$, so we can
apply directly our final result (\ref{dFij}). Using (\ref{C2ZN}) and setting $d=4-\epsilon$, this yields
\begin{equation}
\tilde F_{\textrm{IR}} = \tilde F_{\rm free}-\frac{N(3(g^*_1)^2+(N-1)(g^*_2)^2)\epsilon}{576(4\pi)^3}+\mathcal{O}(\epsilon^4)\,.
\end{equation}
It is straightforward to check that this agrees with the result obtained by the approach of \cite{Giombi:2014xxa}, where
the leading contribution to $\delta\tilde F$ just comes from the integrated 2-point function on the sphere.

Using (\ref{gsZN}), we then get for the three fixed points
\begin{equation}
\begin{aligned}
\tilde F_{O(N)} &= \tilde F_{\rm free} -\frac{N(N+2)\pi}{576(N+8)^2}\epsilon^3+\mathcal{O}(\epsilon^4)\,,\\
\tilde F_{N-{\rm Ising}} = N \tilde F_{\rm Ising} &= \tilde F_{\rm free}-N \frac{\pi}{15552}\epsilon^3+\mathcal{O}(\epsilon^4)\,,\\
\tilde F_{{\rm anisotropic}} &= \tilde F_{\rm free}-\frac{(N+2)(N-1)\pi}{15552N}\epsilon^3+\mathcal{O}(\epsilon^4)\,.
\end{aligned}
\end{equation}
Of course, the leading correction to $\tilde F$ is negative in all cases, consistently with the fact that
we can always flow from the UV free theory to any of the fixed points. It is interesting to
examine the constraints imposed by the Generalized $F$-theorem on the possible flows between them. For instance,
from the above result we see that
\begin{equation}
\begin{aligned}
&\tilde F_{O(N)} > \tilde F_{{\rm anisotropic}}\qquad \mbox{for $N>4$}\,,\\
\end{aligned}
\end{equation}
which is precisely consistent with the fact that for $N>4$ the $O(N)$ fixed point has one relevant operator which triggers
a flow to the anisotropic fixed point, and vice-versa for $N<4$. We also find that
\begin{equation}
\begin{aligned}
&N \tilde F_{\rm Ising} > \tilde F_{O(N)}\qquad \mbox{for $N<10$}\,,\\
\label{N-Ising-to-ON}
\end{aligned}
\end{equation}
which forbids flows from the tensor product of $N$ Ising models to the $O(N)$ fixed point for $N>10$. This is consistent
with the structure of the one-loop RG flows depicted in Fig.  \ref{Flows}, even though the constraint imposed
by $\tilde F$ appears to be slightly weaker. The result (\ref{N-Ising-to-ON}) just follows from the leading order term in $\tilde F$; as discussed
in section \ref{Pade-section}, including higher orders and doing a Pad\'e extrapolation yields (\ref{N-Ising-to-ON-Pa}) in $d=3$. We also see that
\begin{equation}
\begin{aligned}
&N \tilde F_{\rm Ising} > \tilde F_{{\rm anisotropic}}\qquad \mbox{for $N>2$}\,,\\
\end{aligned}
\end{equation}
which is consistent with the flows between $N$-Ising and anisotropic fixed point for $N>4$ (again, the bound
from $\tilde F$ appears to be slightly weaker). We also note that at large $N$ the anisotropic fixed point gets
very close to the $N$-Ising point. Indeed we have
\begin{equation}
N\tilde F_{{\rm Ising}} = \tilde F_{\rm anisotropic}+\frac{\pi \epsilon^3}{15552}+\mathcal{O}(\frac{1}{N})\,.
\end{equation}
The flow from the $N$ decoupled Ising models to the anisotropic fixed point is caused by the
``double-trace" operator $\sum_{i<j}\phi_{i}^2\phi_{j}^2$; this is why the correction to $\tilde F$ is of
order $N^0$.

As done in eq. (\ref{ON-conf-pert}) for the $O(N)$ model, it is interesting to compute the next order correction to $\tilde F$
at the anisotropic fixed point using the conformal perturbation theory approach, under the assumption that the line integral formula (\ref{deltaFMany})
holds at higher orders with ${\cal C}_{2,ij}(g)={\cal C}_{2,ij}+O(g^2)$. The $\beta$ functions computed in conformal perturbation
theory take the form
\begin{equation}
\begin{aligned}
&\beta_1(g_1,g_2) = -\epsilon g_1 +\frac{\Gamma\left(\frac{d-2}{2}\right)}{4\pi^{d/2}(d-2)}\left(9 g_1^2+(N-1)g_2^2\right)
-\frac{51 g_1^3+5  (N-1) g_1 g_2^2+4(N-1) g_2^3}{64 \pi^4}\left(1+\mathcal{O}(\epsilon)\right)\,,\\
&\beta_2(g_1,g_2) =-\epsilon g_2+ \frac{\Gamma\left(\frac{d-2}{2}\right)}{4\pi^{d/2}(d-2)}\left((N+2)g_2^2+6g_1 g_2\right)
-\frac{9(N-1)g_2^3+36 g_1 g_2^2+15 g_1^2 g_2}{64 \pi^4}\left(1+\mathcal{O}(\epsilon)\right)\,.
\label{beta-conf}
\end{aligned}
\end{equation}
Choosing a piecewise straight path from the origin to the fixed point $g_1^*, g_2^*$, we obtain
\begin{equation}
\begin{aligned}
\tilde F_{\textrm{IR}}-\tilde F_{\textrm{UV}} =
\frac{2\pi^{1+d}}{\Gamma\left(1+d\right)}\int_0^{g_2^*} ({\cal C}_2)_{22}\beta_2(0,g_2) dg_2
+\frac{2\pi^{1+d}}{\Gamma\left(1+d\right)}\int_0^{g_1^*} ({\cal C}_2)_{11}\beta_1(g_1,g_2^*) dg_1\,,
\end{aligned}
\end{equation}
where the value of $g_1^*, g_2^*$ to the needed order can be obtained from the $\beta$-functions (\ref{beta-conf}). Carrying out the
integrations, we find
\begin{equation}
\tilde F_{\textrm{IR}}-\tilde F_{\textrm{UV}} =-\frac{\pi(N+2)(N-1)}{15552N}\eps^3-\frac{\pi(N-1)(73N^{3}+36N^{2}+432N-424)}{559872 N^3}\eps^4
+\mathcal{O}(\eps^5)\,.
\end{equation}
This exactly agrees with the result (\ref{Fcubic-ep4}) obtained using the direct dimensional renormalization in curved space, providing a
non-trivial consistency check of our approach.

\subsection{The Gross-Neveu model in $2+\epsilon$ dimensions}

The techniques of conformal perturbation theory may be also applied when the perturbing operator
is weakly irrelevant and one finds a nearby UV fixed point.
As an example, we consider the Gross-Neveu model\cite{Gross:1974jv} in $d=2+\epsilon$:
\begin{equation}
S=\int d^d x\left(\bar\psi_i \gamma^{\mu}\partial_{\mu}\psi^i +\frac{g}{2}(\bar\psi_i \psi^i)^2\right),
\end{equation}
where $\psi^i$ is a collection of $\tilde N$ Dirac fermions (we will denote $N=\tilde N {\rm tr}{\bf 1}$). The perturbing operator $O(x)=\frac{1}{2}(\bar\psi_i \psi^i)^2$ has dimension $\Delta=2(d-1)$ in the
free IR theory. In $d=2+\epsilon$, we have $\Delta=d+\epsilon$, and the operator is weakly irrelevant.

The two- and three-point
functions of the perturbing operator $O(x)$ in the free theory are
\begin{equation}
\begin{aligned}
&\langle O(x)O(y)\rangle = \frac{1}{2} N(N-1)\left(\frac{\Gamma(\frac{d}{2})}{2\pi^{\frac{d}{2}}}\right)^4\frac{1}{|x-y|^{2(d+\epsilon)}},\\
&\langle O(x)O(y)O(z)\rangle = -N(N-1)(N-2)\left(\frac{\Gamma(\frac{d}{2})}{2\pi^{\frac{d}{2}}}\right)^6 \frac{1}{\left(|x-y||y-z||z-x|\right)^{d+\epsilon}}\,.
\end{aligned}
\end{equation}
We can formally apply the result (\ref{betaPert}) with $\epsilon \rightarrow -\epsilon$ and
obtain the beta function
\begin{equation}
\beta =\epsilon g-\frac{(N-2)\Gamma(\frac{d}{2})}{2\pi^{\frac{d}{2}}}g^2+\left(\frac{N-2}{4\pi^2}+\mathcal{O}(\epsilon)\right)g^3+\mathcal{O}(g^4)\,,
\end{equation}
where we have included the subleading term by requiring that it matches the
known beta function of the Gross-Neveu model in $d=2$ \cite{Moshe:2003xn}.

Applying  (\ref{intBeta}), and again assuming that ${\cal C}_2(g)={\cal C}_2+\mathcal{O}(g^2)$, we obtain
\begin{equation}
\begin{aligned}
&\delta \tilde F = \frac{2\pi^{1+d}}{\Gamma\left(1+d\right)}\int_0^{g_*} {\cal C}_2 \beta(g)dg
=\frac{\pi N(N-1)}{48(N-2)^2}\epsilon^3-\frac{\pi N(N-1)(N-3)}{32(N-2)^3}\epsilon^4+\mathcal{O}(\epsilon^5)\,.
\end{aligned}
\end{equation}
As shown in \cite{FGKT-to-appear}, this precisely agrees with the direct field theory calculation in curved space in $d=2+\epsilon$, where
the Euler curvature beta function is set to its zero, following the same logic as outlined in Section \ref{WF-curved}.



\section*{Acknowledgments}
We thank Z. Komargodski, R. Melko, H. Osborn and M. Smolkin for useful discussions.
The work of LF and SG was supported in part by the US NSF under Grant No.~PHY-1318681.
The work of IRK and GT was supported in part by the US NSF under Grant No.~PHY-1314198.

\appendix

\section{General quartic theory on the sphere}
\label{gen-coup}
In this Appendix we consider, following \cite{Jack:1983sk,Jack:1990eb}, a general $\phi^{4}$  theory with the action
\begin{align}
S= \int d^{d}x \left(\frac{1}{2}\big((\partial_{\mu}\phi^{i}_{0})^{2}+ \frac{d-2}{4(d-1)}\mathcal{R}(\phi_{0}^{i})^{2}\big)+\frac{1}{4}\lambda^{0}_{ijkl}\phi_{0}^{i}\phi_{0}^{j}\phi_{0}^{k}\phi_{0}^{l}+\frac{1}{2}\eta_{0}H(\phi_{0}^{i})^{2}+a_{0}W^2+b_{0}E+c_{0}H^{2}\right),
\end{align}
where $\lambda^{0}_{ijkl}$ is a symmetric tensor.  The $\beta$-functions are given by \cite{Jack:1983sk}
\begin{align}
\beta^{\lambda}_{ijkl}=&-\epsilon \lambda_{ijkl} + \frac{6}{(4\pi)^{2}} \beta^{(2)}_{ijkl} +\frac{3}{(4\pi)^{4}}\big(\beta^{(3,1)}_{ijkl}-12 \beta^{(3,2)}_{ijkl}\big)\notag\\
&+\frac{27}{2(4\pi)^{6}} \Big(8 \beta_{ijkl}^{(4,1)}-6 \beta_{ijkl}^{(4,2)}+32 \beta_{ijkl}^{(4,3)}-4\beta_{ijkl}^{(4,4)}-2\beta_{ijkl}^{(4,5)}+8 \zeta (3) \beta_{ijkl}^{(4,6)}-\beta_{ijkl}^{(4,7)}\Big)+...\,, \label{betatens}
\end{align}
where
\begin{align}
&\beta^{(2)}_{ijkl}= \lambda_{abij} \lambda_{abkl}+\dots, \quad \beta^{(3,1)}_{ijkl}= \lambda_{abcd} \lambda_{aijk} \lambda_{lbcd}+\dots, \quad \beta^{(3,2)}_{ijkl} = \lambda_{abij} \lambda_{acdk} \lambda_{bcdl}+\dots \notag\\
 \beta_{ijkl}^{(4,1)} &=\lambda_{abef} \lambda_{cdef} \lambda_{acij} \lambda_{bdkl}+\dots, \quad  \beta_{ijkl}^{(4,2)} =\lambda_{acde} \lambda_{afij} \lambda_{bcde}\lambda_{bfkl}+\dots, \quad
  \beta_{ijkl}^{(4,3)} =\lambda_{abij} \lambda_{acde} \lambda_{befl} \lambda_{cdfk}+\dots,\notag\\
  \beta_{ijkl}^{(4,4)} &=\lambda_{abij} \lambda_{acdk} \lambda_{befl} \lambda_{cdef}+\dots,\quad     \beta_{ijkl}^{(4,5)} =\lambda_{acdi} \lambda_{aefk} \lambda_{bcdj} \lambda_{befl}+\dots,\quad
  \beta_{ijkl}^{(4,6)} =\lambda_{abci} \lambda_{adej} \lambda_{dbfk} \lambda_{fecl}+\dots, \notag\\
    \beta_{ijkl}^{(4,7)} &=\lambda_{abcd} \lambda_{aikl} \lambda_{jbef} \lambda_{cdef}+\dots \label{betas}
\end{align}
and the ellipses mean all inequivalent permutations of $i,j,k,l$\;\footnote{For instance $\beta^{(2)}_{ijkl}=  \lambda_{abij} \lambda_{abkl}+ \lambda_{abik} \lambda_{abjl}+ \lambda_{abil} \lambda_{abjk}$ and  $\beta^{(3,1)}_{ijkl}= \lambda_{abcd} \lambda_{aijk} \lambda_{lbcd}+ \lambda_{abcd} \lambda_{aijl} \lambda_{kbcd}+ \lambda_{abcd} \lambda_{aikl} \lambda_{jbcd}+ \lambda_{abcd} \lambda_{ajkl} \lambda_{ibcd}$.}.
One can recover the usual $O(N)$ theory using the tensor
\begin{align}
 \lambda_{ijkl}= \frac{1}{3} \lambda(\delta_{ij}\delta_{kl}+\delta_{ik}\delta_{jl}+\delta_{il}\delta_{jk})\,.
\end{align}
Using $\beta$-functions (\ref{betatens}) one can find how bare couplings are connected with renormalized ones:
\begin{align}
 \lambda^{0}_{ijkl}=\mu^{\epsilon}\bigg( \lambda_{ijkl}+\frac{6\beta^{(2)}_{ijkl}}{(4\pi)^{2}\epsilon}+...\bigg), \label{bareandren}
\end{align}
due to cumbersome expressions we only write the first term.
For the connected correlators we find
\begin{align}
&G_{2}= 4!\, t^{0}_{2}\,C_{\phi}^{4}I_{2}(2(d-2))\,, \notag\\
&G_{3}=3(4!)^{2}\,t^{0}_{3}\,C_{\phi}^{6}I_{3}(2d-4)\,, \notag\\
&G_{4}=12(4!)^{2}\big(9t^{0}_{41}G_{4}^{(1)}+8t^{0}_{42}G_{4}^{(2)}+36t^{0}_{43}G_{4}^{(3)}\big)\,, \notag\\
&G_{5}=(4!)^{4}\big(60 t^{0}_{51}G_{5}^{(1)}+90 t^{0}_{52}G_{5}^{(2)}+9 t^{0}_{53}G_{5}^{(3)}
+10 t^{0}_{54}G_{5}^{(4)}+180 t^{0}_{55}G_{5}^{(5)}+24 t^{0}_{56}G_{5}^{(6)} \big)\,,
\label{general-Gs}
\end{align}
where the tensor structures are easily read off from the diagrams in figure \ref{diagsPhi4}:
\begin{align}
&t_{2}=\lambda_{ijkl}\lambda_{ijkl}, \quad t_{3}=\lambda_{ijkl}\lambda_{ijmn}\lambda_{klmn}, \notag\\
&t_{41}=\lambda_{ijkl}\lambda_{ijmn}\lambda_{mnpr}\lambda_{prkl},\quad t_{42}=\lambda_{ijkl}\lambda_{lmnp}\lambda_{mnpr}\lambda_{ijkr},\quad t_{43}=\lambda_{ijkl}\lambda_{kpmn}\lambda_{mnlr}\lambda_{ijpr}, \notag\\
&t_{51}=\lambda_{ijkl}\lambda_{klmn}\lambda_{ijmp}\lambda_{prst}\lambda_{nrst},\quad t_{52}=\lambda_{ijkl}\lambda_{klmn}\lambda_{prsn}\lambda_{prmt}\lambda_{ijst},\quad t_{53}=\lambda_{ijkl}\lambda_{klmn}\lambda_{mnpr}\lambda_{prst}\lambda_{ijst}, \notag\\
&t_{54}=\lambda_{ijkl}\lambda_{ijkm}\lambda_{lmnp}\lambda_{prst}\lambda_{nrst},\quad t_{55}=\lambda_{ijkl}\lambda_{lmnp}\lambda_{pnkr}\lambda_{jrst}\lambda_{imst},\quad t_{56}=\lambda_{ijkl}\lambda_{lmnp}\lambda_{pksr}\lambda_{jrnt}\lambda_{imst}.
\end{align}
Using (\ref{betas}) and (\ref{bareandren}) we find
\begin{align}
t_{2}^{0}=& \mu^{2\epsilon}\bigg(t_{2}+\frac{36t_{3}}{(4\pi)^{2}\epsilon}+\Big(\frac{12(t_{42}-18 t_{43})}{(4\pi)^{4}\epsilon}+\frac{324(t_{41}+2t_{43})}{(4\pi)^{4}\epsilon^{2}}\Big)+\Big(-\frac{18 (11 t_{51}+12 t_{52}-168 t_{55}-96 t_{56} \zeta (3))}{(4 \pi )^6 \epsilon }\notag\\
&\qquad+\frac{720 (t_{51}-6 t_{52}-122 t_{55})}{(4 \pi )^6 \epsilon ^2}+\frac{2592 (4 t_{52}+t_{53}+4 t_{55})}{(4 \pi )^6 \epsilon ^3}\Big)+...\bigg)\,, \notag\\
t_{3}^{0}=& \mu^{3\epsilon}\bigg(t_{3}+\frac{18(t_{41}+2t_{43})}{(4\pi)^{2}\epsilon}+\Big(\frac{18 (t_{51}-6 t_{52}-12 t_{55})}{(4 \pi )^4 \epsilon }+\frac{216 (4 t_{52}+t_{53}+4 t_{55})}{(4 \pi )^4 \epsilon ^2}\Big)+...\bigg)\,, \notag\\
t_{41}^{0}=&\mu^{4\epsilon}\bigg(t_{41}+\frac{24 (2 t_{52}+t_{53})}{(4 \pi )^2 \epsilon }+...\bigg)\,, \quad t_{42}^{0}=\mu^{4\epsilon}\bigg(t_{42}+\frac{72 t_{51}}{(4 \pi )^2 \epsilon }+...\bigg)\,, \quad t_{43}^{0}=\mu^{4\epsilon}\bigg(t_{43}+\frac{24 (t_{52}+2 t_{55})}{(4 \pi )^2 \epsilon }+...\bigg)\,. \label{tbaretren}
\end{align}
Demanding that the expression for the sphere free energy
\begin{align}
F-F_{\textrm{free}} = -\frac{G_{2} }{2!\cdot 4^{2}}+ \frac{G_{3}}{3!\cdot 4^{3}} - \frac{G_{4}}{4!\cdot 4^{4}} + \frac{G_{5}}{5!\cdot 4^{5}} +\int d^{d}x\sqrt{g} (b_{0}E+c_{0}H^{2})
\label{F-general}
\end{align}
have no poles in $\epsilon$ cannot fix the counter terms completely, because $E$ and $H^2$ are
proportional to each other on a round sphere. However, using also the result for $c_{51}$ from \cite{Jack:1990eb},
we find
\begin{align}
&b_{0}=\mu ^{-\epsilon } \left(b+\frac{b_{01}}{\epsilon}+\frac{t_{42}-18 t_{43}}{80 (4 \pi )^{10}\epsilon}+\Big(-\frac{11 t_{51}+51 t_{52}-186 t_{55}-144 t_{56} \zeta (3)}{40 (4 \pi )^{12} \epsilon }+\frac{3 (t_{51}-6 t_{52}-12 t_{55})}{20 (4 \pi )^{12} \epsilon ^2}\Big)+...\right), \notag\\
&c_{0}=\mu ^{-\epsilon } \left(c-\frac{3 (t_{51}-6 t_{52}-12 t_{55})}{40 (4 \pi )^{12}\epsilon}+...\right)\,,
\label{curv-ren-general}
\end{align}
where $b_{01}=-\frac{N}{360(4\pi)^2}$ is the free field contribution determined by (\ref{Ffree-div}), which just depends on the total number of fields.
For the curvature $\beta$-functions we have therefore
\begin{align}
&\beta_{b}= \epsilon b + b_{01}+\frac{t_{42}-18 t_{43}}{16 (4 \pi )^{10}}-\frac{3(11 t_{51}+51 t_{52}-186 t_{55}-144 t_{56} \zeta (3))}{20 (4 \pi )^{12}  }\,, \notag\\
&\beta_{c}=\epsilon c -\frac{9 (t_{51}-6 t_{52}-12 t_{55})}{20 (4 \pi )^{12}}\,.
\end{align}
All these results are in agreement with the results for the $O(N)$ model and the first terms in $\beta_{b}$ and $\beta_{c}$ agree with \cite{Jack:1983sk,Jack:1990eb}. The
term of order $\lambda^5$ in $\beta_b$ is, as far as we know, a new result.

\subsection{The $O(N)$ model with anisotropy}
\label{cubic-App}
As an interesting special case, let us consider the $O(N)$ model with anisotropy, whose action is given in (\ref{cubicS}).
In this case our $\lambda_{ijkl}$ tensor is given by
\begin{align}
\lambda_{ijkl} =(g_{1}-g_{2})\delta_{ijkl} +\frac{g_{2}}{3}(\delta_{ij}\delta_{kl}+\delta_{ik}\delta_{jl}+\delta_{il}\delta_{jk})\,,
\end{align}
where $\delta_{ijkl}=1$  if $i=j=k=l$ and $\delta_{ijkl}=0$ in  another case.

The relation between bare and renormalized couplings reads
\begin{align}
g_{1,0} =& \mu^\eps \bigg(g_1+\frac{9g_1^2+(N-1)g_2^2}{8\pi^2\eps}+\Big(\frac{81g_1^3+15(N-1)g_1g_2^2+(N^2+N-2)g_2^3}{64\pi^4\eps^2} \notag\\
&\qquad-\frac{51g_1^3+5(N-1)g_1g_2^2+4(N-1)g_2^3}{128\pi^4\eps}\Big)\notag\\
&\qquad+\frac{1}{512\pi^6\eps^3}\Big(729g_1^4+ 186(N-1)g_1^2 g_2^2 +16(N-1)(N+2)g_1 g_2^3+(N-1)(N^2+9N-1)g_2^4)\Big)\notag\\
&\qquad+\frac{1}{3072\pi^6\eps^2}\Big(-3213 g_1^4-498(N-1)g_1^2 g_2^2-10(N-1)(N+38)g_1 g_2^3-(N-1)(53N-17)g_2^4\Big)\notag\\
&\qquad+\frac{1}{12288\pi^6\eps}\Big(27(145+96\zeta(3))g_1^4+162(N-1)g_1^2g_2^2-2(N-1)(13N-430-192\zeta(3))g_1 g_2^3 \notag\\
&\qquad\qquad\qquad\quad+(N-1)(59N-67+96\zeta(3))g_2^4 \Big)+...\bigg), \notag\\
g_{2,0}=& \mu^\eps \bigg( g_2+\frac{6g_1g_2+(N+2)g_2^2}{8\pi^2\eps}+\Big(\frac{45 g_1^2g_2+9(N+2)g_1g_2^2+(N^2+7N+1)g_2^3}{64\pi^4\eps^2}\notag\\
&\qquad-\frac{15g_1^2g_2+36g_1g_2^2+9(N-1)g_2^3}{128\pi^4 \eps}\Big)\notag\\
&\qquad+\frac{1}{512\pi^6\eps^3}\Big(360g_1^3g_2+78(N+2)g_1^2 g_2^2+12N(N+8)g_1 g_2^3+(N+2)(N^2+10N-2)g_2^4\Big)\notag\\
&\qquad+\frac{1}{1024\pi^6\eps^2}\Big(-384g_1^3g_2 -(25N+446)g_1^2g_2^2 - 24(8N+1)g_1 g_2^3-7(N-1)(3N+10)g_2^4\Big)\notag\\
&\qquad+\frac{1}{12288\pi^6\eps}\Big(1512g_1^3 g_2 + 9(N+138+192\zeta(3)g_1^2 g_2^2)+24(19N+37+48\zeta(3))g_1 g_2^3 \notag\\
&\qquad\qquad\qquad+(33N^2+457N-682-768\zeta(3)+480 N \zeta(3))g_2^4\Big)+...\bigg).
\end{align}
and for the $\beta$-functions we find \cite{Kleinert:2001ax}
\begin{align}
\beta_{1}=& -\epsilon g_{1}+ \frac{9 g_{1}^2+(N-1)g_{2}^2 }{8 \pi ^2}-\frac{51 g_{1}^3+5(N-1) g_{1} g_{2}^2 +4  (N-1)g_{2}^3}{64 \pi ^4}  + \frac{1}{4096\pi^6}\Big(27(145+96\zeta(3))g_1^4\notag\\
&+162(N-1)g_1^2g_2^2-2(N-1)(13N-430-192\zeta(3))g_1g_2^3+(N-1)(59N-67+96\zeta(3))g_2^4\Big)\,,\notag\\
\beta_{2}=&-\epsilon g_{2}+\frac{6 g_{1} g_{2}+ (N+2)g_{2}^2}{8 \pi ^2}-\frac{15 g_{1}^2 g_{2}+36 g_{1} g_{2}^2+9 (N-1) g_{2}^3}{64 \pi ^4}+ \frac{g_{2}}{4096\pi^6}\Big(9(N+138+192\zeta(3))g_1^2g_2\notag\\
&+1512g_1^3+24(19N+37+48\zeta(3))g_1g_2^2+(33N^2+457N-682+96(5N-8)\zeta(3))g_2^3\Big)\,.
\label{beta-aniso}
\end{align}
The tensor structures defined in the previous section read explicitly
\begin{align}
&t_{2}=\frac{1}{3} N \big(3 g_{1}^2+(N-1)g_{2}^2 \big),\notag\\
&t_{3}=\frac{1}{3^{3}} N \big(27 g_{1}^3+9 (N-1)g_{1} g_{2}^2 + (N-1) (N+2)g_{2}^3\big), \notag\\
&t_{41}=\frac{N}{3^{4}} \big(81 g_{1}^4+54(N-1) g_{1}^2 g_{2}^2 +36(N-2) (N-1) g_{1} g_{2}^3 +(N-1) (N^2-3 N+11)g_{2}^4 \big), \notag\\
&t_{42}=\frac{N}{3^2} \big(9 g_{1}^4+6 (N-1) g_{1}^2 g_{2}^2+ (N-1)^2 g_{2}^4\big),\notag\\
&t_{43}=\frac{N}{3^4}\big(81 g_{1}^4+18(N-1) g_{1}^2 g_{2}^2 +24 (N-1)g_{1} g_{2}^3 +5 (N-1)^2g_{2}^4 \big)
\end{align}
and
\begin{align}
t_{51}=&\frac{N}{3^4} \Big(3 g_{1}^2+(N-1)g_{2}^2 ) (27 g_{1}^3+9(N-1) g_{1} g_{2}^2 + (N^2+N-2)g_{2}^3\Big), \notag\\
t_{52}=&\frac{N}{3^5}\Big(243g_{1}^5+108 (N-1) g_{1}^3 g_{2}^2 +9  (N-1) (N+4) g_{1}^2 g_{2}^3+3  (9 N-7) (N-1) g_{1} g_{2}^4\notag\\
&\qquad+(N-1) (3 N^2-5 N+8)g_{2}^5 \Big),\notag\\
t_{53}=&\frac{N}{3^5} \Big(243 g_{1}^5+270  (N-1) g_{1}^3 g_{2}^2+90(N-2) (N-1) g_{1}^2 g_{2}^3 +15  (N-1) (N^2-3 N+3) g_{1} g_{2}^4\notag\\
&\qquad+(N-1) (N^3-4 N^2+6 N+12)g_{2}^5 \Big), \notag\\
t_{54}=&\frac{N}{3^3}\big(3 g_{1}+(N-1)g_{2} \big) \big(3 g_{1}^2+(N-1)g_{2}^2 \big)^2, \notag\\
t_{55}=&\frac{N}{3^5}\Big(243 g_{1}^5+54 (N-1) g_{1}^3 g_{2}^2+72(N-1) g_{1}^2 g_{2}^3 +3 (N-1) (5 N+3) g_{1} g_{2}^4 + (N-1) (N^2+8 N-12)g_{2}^5\Big), \notag\\
t_{56}=&\frac{N}{3^4}\Big(81 g_{1}^5+30 (N-1)g_{1}^2 g_{2}^3 +15(N-1) g_{1} g_{2}^4 +  (N-1)(5N-8)g_{2}^5\Big).
\end{align}
Writing (note that the coupling constants are absorbed in the coefficients $b_{ij}, c_{ij}$ below)
\begin{equation}
\begin{aligned}
&b_0 =\mu^{-\epsilon}\left(b+\frac{b_{01}}{\epsilon}+\frac{b_{41}}{\epsilon}+\Big(\frac{b_{52}}{\epsilon^2}+\frac{b_{51}}{\epsilon}\Big)+\ldots \right),\\
&c_0 = \mu^{-\epsilon}\left(c+\frac{c_{51}}{\epsilon}+\ldots\right)
\end{aligned}
\end{equation}
we find, using eq.~(\ref{curv-ren-general}):
\begin{align}
b_{41}=&-\frac{N \left(51 g_{1}^4+10(N-1) g_{1}^2 g_{2}^2 +16  (N-1)g_{1} g_{2}^3+3 (N-1)^2 g_{2}^4\right)}{240 (4\pi) ^{10}}\,,\notag\\
b_{51}=&\frac{N}{540 (4\pi)^{12}} \Big(54 g_{1}^5 (36 \zeta (3)+31)+153 (N-1) g_{1}^3 g_{2}^2 -(N-1) (31 N-631-720\zeta(3))g_{1}^2 g_{2}^3 \notag\\
&+(N-1) (62 N+360 \zeta (3)+169)g_{1} g_{2}^4 + (N-1) (24 (5 N-8) \zeta (3)+95 N-143)g_{2}^5\Big)\,, \notag\\
c_{51}=&\frac{N}{360 (4\pi)^{12}} \Big(459 g_{1}^5+126(N-1) g_{1}^3 g_{2}^2 + (N-1) (5 N+118)g_{1}^2 g_{2}^3+(N-1) (35 N+1)g_{1} g_{2}^4 \notag\\
&+ (N-1)^2 (3 N+10)g_{2}^5\Big)\,.
\end{align}
The curvature beta functions are then given by
\begin{equation}
\beta_b = \epsilon b+b_{01}+5b_{41}+6 b_{51}+\ldots \,,\qquad \beta_c = \epsilon c+6 c_{51}+\ldots\,.
\end{equation}
From these we can obtain the values $b_*$ and $c_*$ at the IR fixed point $\beta_b=\beta_c =\beta_1=\beta_2=0$,
as a function of the couplings $g_1^*$, $g_2^*$. At the anisotropic fixed
point, the latter are given by
\begin{align}
g_{1}^*=&\frac{8\pi^2(N-1)}{9N}\eps + \frac{8(106-159N+36N^2+17N^3)\pi^2}{243N^3}\eps^2+\frac{\pi^2}{6561N^5}\Big(N\big(418304-299734N+54649N^2\notag\\
&+5848N^3+709N^4-2592(N-1)(N^3+3N^2+N-14)\zeta(3)\big)-179776\Big)\eps^3+\mathcal{O}(\eps^4) \notag\\
g_{2}^*=&\frac{8\pi^2}{3N}\eps-\frac{8\pi^2(106-125N+19N^2)}{81N^3}\eps^2-\frac{\pi^2}{2187N^5} \Big(N\big(360640-229974N+41971N^2\notag\\
&+1955N^3-2592(2N^{3}-6N^{2}-7N+14)\zeta(3))\big)-179776\Big)\eps^3+\mathcal{O}(\eps^4)\,.
\label{gstar-cubic}
\end{align}
Specializing (\ref{general-Gs}) to the present case and using (\ref{F-general}), we find that the free energy at the anisotropic fixed point is:
\begin{align}
&\tilde{F}_{\textrm{cubic}}-N \tilde{F}_{\textrm{free}} =-\frac{(N+2)(N-1)\pi}{15552N}\eps^3-\frac{(N-1)(73N^{3}+36N^{2}+432N-424)\pi}{559872N^3}\eps^4 \notag \\
&~~+ \frac{\pi}{302330880N^5}\Big(-53621N^6+14259N^5+224610N^4-2161280N^3+5542944N^2-5364672N+1797760\notag\\
&\qquad\quad+810\pi^2N^4(N^2+N-2)+10368N(N-1)(N^4-4N^3+16N^2+6N-28)\zeta(3)\Big)\eps^5+\ldots\,.
\label{Fcubic-ep4}
\end{align}
We notice that at
\begin{align}
N=N_{\textrm{crit}}=4-2\eps-\Big(\frac{5}{12}-\frac{5\zeta(3)}{2}\Big)\eps^2+\mathcal{O}(\eps^3)
\end{align}
we find $F_{O(N)}=F_{\textrm{cubic}}$, where $F_{O(N)}$ is given in (\ref{tFON}). This is as expected since for this value of $N$ the couplings
at the $O(N)$ and anisotropic point coincide (compare (\ref{lamstar}) and (\ref{gstar-cubic})).

\section{Feynman Integrals on the Sphere}
\label{Integ-App}
\begin{figure}[h!]
                \centering
                \includegraphics[width=7.5cm]{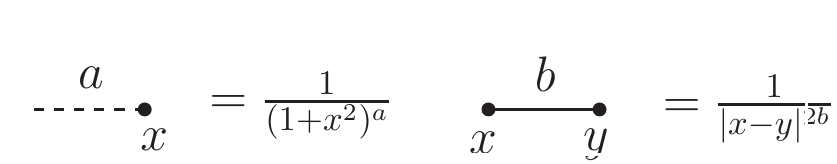}
                \caption{External line and propogator  for the integrals on a sphere.}
                \label{extprop}
\end{figure}

In this appendix we describe a method for computing Feynman integrals on a sphere. We start from the well known
integrals, which can be calculated explicitly \cite{Drummond:1977dn, Cardy:1988cwa, Klebanov:2011gs}:
\begin{align}
&I_{2}(\Delta)=\int \frac{d^{d}xd^{d}y\sqrt{g_{x}}\sqrt{g_{y}}}{s(x,y)^{2\Delta}}= (2 R)^{2 (d-\Delta )}\frac{2^{1-d} \pi ^{d+\frac{1}{2}} \Gamma \left(\frac{d}{2}-\Delta \right)}{\Gamma \left(\frac{d+1}{2}\right) \Gamma (d-\Delta )}\,, \notag\\
& I_{3}(\Delta) =\int \frac{d^{d}xd^{d}yd^{d}z\sqrt{g_{x}}\sqrt{g_{y}}\sqrt{g_{z}}}{[s(x,y)s(y,z)s(z,x)]^{\Delta}}= R^{3(d-\Delta)} \frac{8\pi^{\frac{3(1+d)}{2}}\Gamma(d-\frac{3\Delta}{2})}{\Gamma(d)\Gamma(\frac{1+d-\Delta}{2})^{3}}\,.
\end{align}
Working in stereographic coordinates
all integrals on a sphere have the general form\footnote{In stereographical coordintates the chordal distance is $s(x,y) = \frac{2R|x-y|}{(1+x^{2})^{1/2}(1+y^{2})^{1/2}}$ and the metric is $ds^2= \frac{4R^2 dx^{\mu}dx^{\mu}}{(1+x^2)^2}$ thus  $\int d^{d}x \sqrt{g_{x} } = \int d^{d}x \frac{(2R)^{d}}{(1+|x|^{2})^{d}}$.}
\begin{align}
I=(2R)^{nd-2\sum_{k<l}^{n}b_{kl}}\int \prod_{i=1}^{n}\frac{d^{d}x_{i}}{(1+x_{i}^{2})^{a_{i}}}\frac{1}{\prod_{k<l}^{n}|x_{k}-x_{l}|^{2b_{kl}}}\,.
\end{align}
Using rotational invariance we can always put one point to zero  and multiply our integral by the volume of the sphere
(one can always choose the point which simplifies the subsequent calculations). Then making iversion  $x_{i}^{\mu}\to x^{\mu}_{i}/|x_{i}|^{2}$ we obtain (we choose $x_{n}=0$)
\begin{align}
I=(2R)^{nd-2\sum_{k<l}^{n}b_{kl}}\frac{2^{1-d}\pi^{\frac{d+1}{2}}}{\Gamma(\frac{d+1}{2})}\int \prod_{i=1}^{n-1}\frac{d^{d}x_{i}}{(1+x_{i}^{2})^{a_{i}}}\frac{1}{\prod_{k<l}^{n-1}|x_{k}-x_{l}|^{2b_{kl}}}\,.
\end{align}
Thus we can always reduce the number of integrals by one. We want  to represent each  integral $I$ in the Mellin-Barnes form. After this is done we can use special Mathematica packages\cite{Czakon:2005rk, Smirnov:2009up}  to calculate the value of the integral as a series in $\epsilon$ \cite{Smirnov:2012gma}.
All integrals $I$ on a sphere can be represented diagramatically with the help of  the external line and propagator (Figure \ref{extprop}).

\begin{figure}[h!]
                \centering
                \includegraphics[width=15cm]{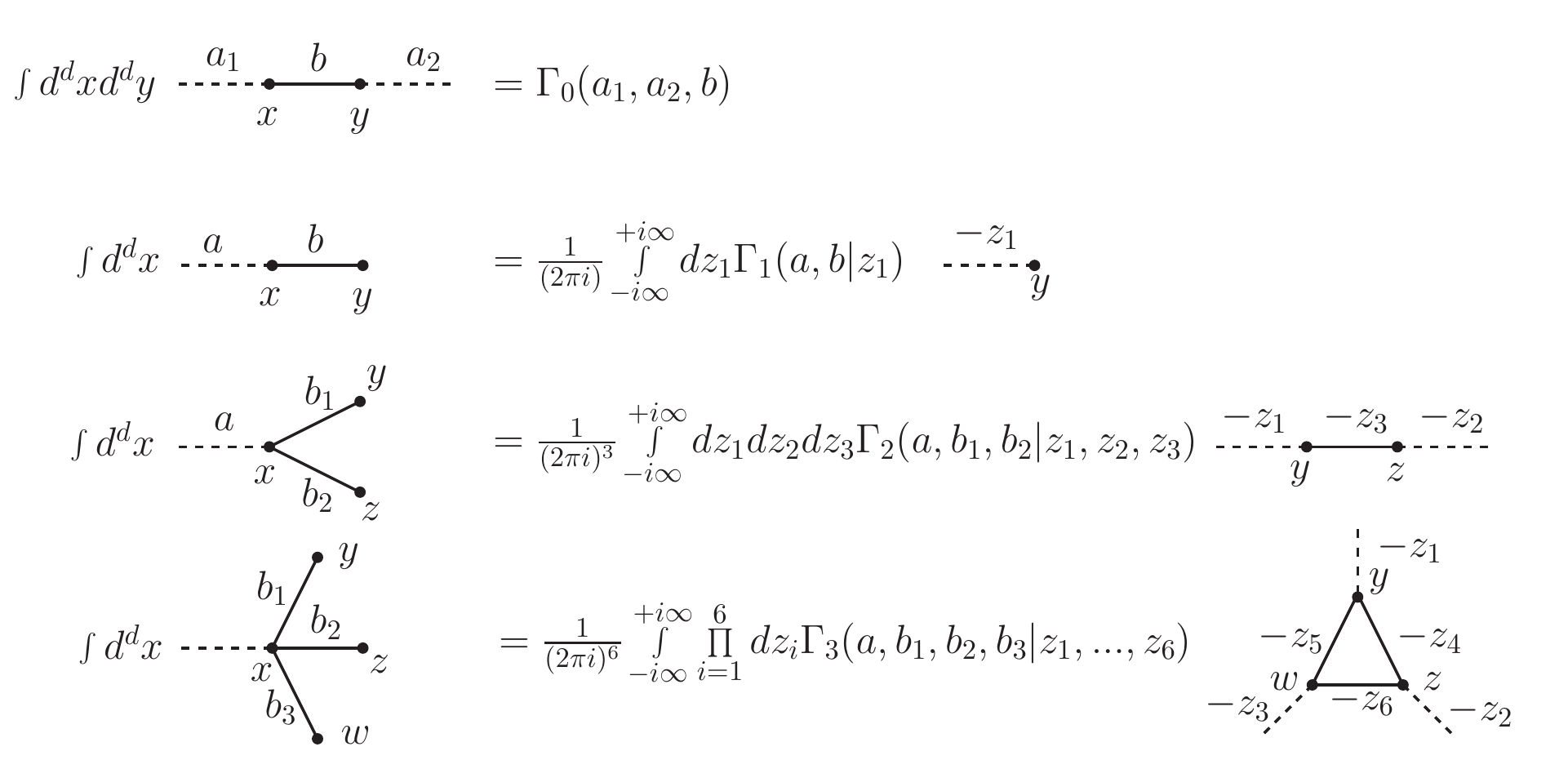}
                                  \caption{Basic relations for rewriting Feynman integrals on a sphere in the Mellin-Barnes form.}
\label{Gammarelations}
\end{figure}

 \noindent Using Feynman parameters and the Mellin-Barnes formula, \footnote{Here we mean the formula $\frac{1}{(X_{1}+X_{2}+...+X_{n})^{\lambda}}=\frac{1}{\Gamma(\lambda)}\frac{1}{(2\pi i)^{n-1}}\int_{-i\infty}^{+i\infty}...\int_{-i\infty}^{+i\infty} dz_{1}...dz_{n-1}\prod_{i=1}^{n-1}X_{i}^{z_{i}}X_{n}^{-\lambda-z_{1}-...-z_{n-1}}\Gamma(\lambda+z_{1}+...+z_{n-1})\prod_{i=1}^{n-1}\Gamma(-z_{i})$.} we  derive  the following  basic relations depicted in Figure \ref{Gammarelations}.

\begin{figure}[h!]
                \centering
                \includegraphics[width=3cm]{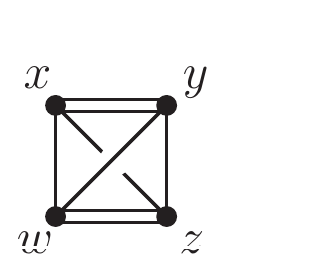}
                  \caption{Diagram $G_{4}^{(3)}$. Each line represents the sphere propogator $\langle \phi(x)\phi(y)\rangle = C_{\phi}/s(x,y)^{d-2}$. Symmetry factor is not included in this graph.}
\label{diagG43}
\end{figure}

\bigskip
\noindent where
\begin{align}
\Gamma_{0}(a_{1},a_{2},b)&=\frac{\pi^{d}}{\Gamma(\frac{d}{2})} \frac{\Gamma(\frac{d}{2}-b)\Gamma(a_{1}+b-\frac{d}{2})\Gamma(a_{2}+b-\frac{d}{2})\Gamma(a_{1}+a_{2}+b-d)}{\Gamma(a_{1})\Gamma(a_{2})\Gamma(a_{1}+a_{2}+2b-d)}\,, \notag\\
\Gamma_{1}(a,b|z_{1})&= \frac{\pi^{d/2}\Gamma(-z_{1})\Gamma(a+b-\frac{d}{2}+z_{1})\Gamma(b+z_{1})\Gamma(d-a-2b-z_{1})}{\Gamma(a)\Gamma(b)\Gamma(d-a-b)}\,,\notag\\
\Gamma_{2}(a,b_{1},b_{2}|z_{1},z_{2},z_{3})&=\frac{\pi^{d/2} \prod\limits_{i=1}^{3}\Gamma(-z_{i})\Gamma(a+b_{1}+b_{2}-\frac{d}{2}+\sum_{i=1}^{3}z_{i}) }{\Gamma(a)\Gamma(b_{1})\Gamma(b_{2})\Gamma(d-a-b_{1}-b_{2})} \Gamma(b_{1}+z_{1}+z_{3})\Gamma(b_{2}+z_{2}+z_{3})
  \notag\\
&\qquad\times  \Gamma(d-a-2b_{1}-2b_{2}-z_{1}-z_{2}-2z_{3})\,, \notag\\
\Gamma_{3}(a,b_{1},b_{2},b_{3}|z_{1},...,z_{6})&=\frac{\pi^{\frac{d}{2}}\prod\limits_{i=1}^{6}\Gamma(-z_{i})\Gamma(a+b_{1}+b_{2}+b_{3}-\frac{d}{2}+\sum_{i=1}^{6}z_{i})}{\Gamma(a)\Gamma(b_{1})\Gamma(b_{2})\Gamma(b_{3})\Gamma(d-a-b_{1}-b_{2}-b_{3})} \Gamma(b_{1}+z_{1}+z_{4}+z_{5})
\notag\\
\times
\Gamma(b_{2}+z_{2}+z_{4}+z_{6})&\Gamma(b_{3}+z_{3}+z_{5}+z_{6}) \Gamma(d-a-2b_{1}-2b_{2}-2b_{3}-z_{1}-z_{2}-z_{3}-2z_{4}-2z_{5}-2z_{6})\,. \label{Gammablocks}
\end{align}
Using these relations one can easily write any integral on a sphere in the Mellin-Barnes form.
For example let us consider the diagram $G_{4}^{(3)}$, which contributes to the free energy of $\phi^{4}$-theory at $\lambda^{4}$ order. This diagram is depicted on Figure \ref{diagG43}.
In this graph each line represents the propagator $C_{\phi}/s(x,y)^{d-2}$ and we don't include the symmetry factor, so the integral reads
\begin{align}
G_{4}^{(3)}=C_{\phi}^{8}\int \frac{d^{d}xd^{d}yd^{d}zd^{d}w \sqrt{g_{x}}\sqrt{g_{y}}\sqrt{g_{z}}\sqrt{g_{w}}}{s(x,y)^{2(d-2)}s(x,z)^{d-2}s(x,w)^{d-2}s(y,z)^{d-2}s(y,w)^{d-2}s(z,w)^{2(d-2)}}\,.
\end{align}

\begin{figure}[h!]
                \centering
                \includegraphics[width=8cm]{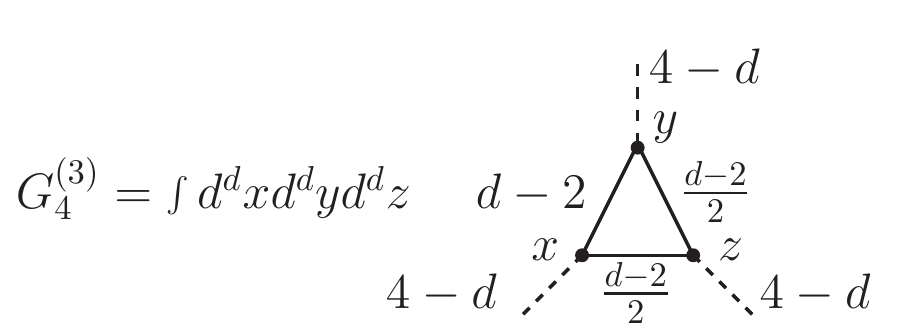}
                \caption{Diagramatic representation for the $G_{4}^{(3)}$ sphere integral.}
                \label{diagG43red}
\end{figure}

\noindent Working in stereographical coordintates and using rotational invarinace to set $w=0$, we arrive at the following integral\footnote{We also need to make inversion to get rid of terms $|x|^{d-2}|y|^{d-2}|z|^{2(d-2)}$ in the denominator.}:
\begin{align}
G_{4}^{(3)}=C_{\phi}^{8}  (2R)^{4(4-d)}\frac{2^{1-d}\pi^{\frac{d+1}{2}}}{\Gamma(\frac{d+1}{2})} \int \frac{d^{d}xd^{d}zd^{d}y}{\big((1+|x|^{2})(1+|z|^{2})(1+|y|^{2})\big)^{4-d}}
\frac{1}{|x-y|^{2(d-2)}|x-z|^{d-2}|y-z|^{d-2}}\,.
\end{align}
Diagramatically this integral is represented (up to some prefactor) in Figure \ref{diagG43red}.

\noindent Using the relations represented in Figure \ref{Gammarelations} we easily get
\begin{align}
G_{4}^{(3)}=&C_{\phi}^{8}  (2R)^{4(4-d)}\frac{2^{1-d}\pi^{\frac{d+1}{2}}}{\Gamma(\frac{d+1}{2})} \notag\\
&\times \frac{1}{(2\pi i)^{3}}\int_{-i\infty}^{i\infty} dz_{1}dz_{2}dz_{3} \Gamma_{2}(4-d,d-2,\frac{d-2}{2}|z_{1},z_{2},z_{3})\Gamma_{0}(4-d-z_{1},4-d-z_{2},\frac{d-2}{2}-z_{3})\,,
\end{align}
where the functions  $\Gamma_{2}$ and $\Gamma_{0}$ are given in (\ref{Gammablocks}). In practice one should try to use the relations in Figure  \ref{Gammarelations} in a way which gives the least number of integrations over parameters $z_{i}$ in the Mellin-Barnes representation.

\bibliographystyle{ssg}
\bibliography{GenF}

\end{document}